\documentclass{aa}  

\usepackage{graphicx}
\usepackage{graphicx}
\usepackage{lscape}
\usepackage{longtable}
\usepackage{natbib}
\usepackage{color}
\usepackage{array} %%%%%%%%%%%%%%%%%%%%%%%%%%%%%%%%%%%%%%%%
\usepackage{txfonts}
\usepackage{color}
\begin{document}

   \title{Eclipsing binaries and fast rotators in the {\it Kepler} sample}

   \subtitle{Characterization via radial velocity analysis from Calar Alto\thanks{Based on observations collected at the German-Spanish Astronomical Center, Calar Alto, jointly operated by the Max- Planck-Institut fur Astronomie (Heidelberg) and the Instituto de Astrof\'isica de Andaluc\'ia (IAA-CSIC, Granada). } }

   \author{J. Lillo-Box\inst{1}, D. Barrado\inst{1}, L. Mancini\inst{2}, Th. Henning\inst{2}, P. Figueira\inst{3,4}, S. Ciceri\inst{2}, N. Santos\inst{3,4,5}
          }

%1
  \institute{Depto. de Astrof\'isica, Centro de Astrobiolog\'ia (CSIC-INTA), ESAC campus 28691 Villanueva de la Ca\~nada (Madrid), Spain\\
              \email{Jorge.Lillo@cab.inta-csic.es}\and
%2
Max Planck Institute for Astronomy, K\"onigstuhl 17, 69117 Heidelberg, Germany  \and %\and
%3
Centro de Astrof\'{i}sica, Universidade do Porto, Rua das Estrelas, 4150-762 Porto, Portugal\and
%5
Instituto de Astrof\' isica e Ci\^encias do Espa\c{c}o, Universidade do Porto, CAUP, Rua das Estrelas, PT4150-762 Porto, Portugal\and
 %4
Departamento de F\'{i}sica e Astronomia, Faculdade de Ci\^{e}ncias, Universidade do Porto, Portugal 
            }

   \date{Accepted by A\&A on January 20th, 2015.}

% \abstract{}{}{}{}{} 
% 5 {} token are mandatory
 
  \abstract
   %{} leave it empty if necessary  
   {The {\it Kepler} mission has provided  high-accurate photometric data in a long timespan for more than two hundred thousands stars, looking for planetary transits. Among the thousands of candidates detected, { the planetary nature} of around 15\% has been { established} or validated by different techniques. But additional data is needed to { characterize} the rest of the candidates and reject other possible configurations.}
   {We started a follow-up program to validate, confirm, and characterize some of the planet candidates. In this paper we present the radial velocity analysis of those presenting large variations, compatible with being eclipsing binaries. We also study those showing large rotational velocities, which prevents us from obtaining the necessary precision to detect planetary-like objects.}
   {We present new radial velocity results for 13 {\it Kepler} objects of interest (KOIs) obtained with the CAFE spectrograph at the Calar Alto Observatory, and analyze their high-spatial resolution (lucky) images obtained with AstraLux and the {\it Kepler} light curves of some interesting cases. }
   {We have found five spectroscopic and eclipsing binaries (group A). Among them, the case of KOI-3853 is of particular interest. This system is a new example of the so-called \emph{heartbeat stars}, showing dynamic tidal distortions in the {\it Kepler} light curve. We have also detected  duration and depth variations of the eclipse. We suggest possible scenarios to explain such effect, including the presence of a third substellar body possibly detected in our radial velocity analysis. We also provide upper mass limits to the transiting companions of other six KOIs with large rotational velocities (group B). This property prevents the radial velocity method to obtain the necessary precision to detect planetary-like masses. Finally, we analyze the large radial velocity variations of other two KOIs, incompatible with the presence of planetary-mass objects (group C). { These objects are likely to be stellar binaries.} However, a longer timespan is needed to complete their characterization.}
   {}

               
\titlerunning{Eclipsing binaries and fast rotators in the {\it Kepler} sample}
\authorrunning{Lillo-Box et al. }
   \maketitle
%

%%%%%%%%%%%%%%%%%%%%%%%%%%%%%%%%%%%
%             		INTRODUCTION
%%%%%%%%%%%%%%%%%%%%%%%%%%%%%%%%%%%

\section{Introduction}

The need of ground-based follow-up observations of the planet candidates provided by the {\it Kepler} mission \citep{borucki10} was known even prior to its launch on March 2009. This need comes from two sides. First, high spatial resolution is required owing to the large pixel size ($\sim$ 4 arcsec) and the typical photometric apertures (6-10 arcsec) used by {\it Kepler}, which permits the contamination of light-curve by close sources, micking planet-like transits. Second, radial velocity (RV) observations of the isolated candidates are crucial to { identify} or reject their planetary nature. The Community Follow-up Observing Program (CFOP\footnote{https://cfop.ipac.caltech.edu/home/}) serves as a communication tool between the different international groups working on this. The false positive rate in the {\it Kepler} sample of planet candidates has been investigated by several authors \citep[e.g., ][]{morton11, santerne12, fressin13, coughlin14}, and more than 3000 {\it Kepler} Objects of Interest (KOIs) have been rejected so far by different groups using different techniques\footnote{ According to the CFOP and the NASA Exoplanet Archive \citep[http://exoplanetarchive.ipac.caltech.edu, ][]{akeson13}, among the 7348 identified KOIs to date, 3170 have been labelled as "False positives".}.

In 2011, we started a ground-based follow-up program with both high-resolution (lucky) imaging with AstraLux  \citep{lillo-box12,lillo-box14b} to detect possible close companions and high-resolution spectroscopy with CAFE\footnote{Calar Alto Fiber-fed Echelle spectrograph} \citep{aceituno13} to { characterize} some of the {\it Kepler} candidates via RV studies, both instruments installed at the 2.2m telescope in the Calar Alto Observatory.  

In this paper, we present the results of part of our high-resolution spectroscopic survey of isolated planet candidates in the {\it Kepler} sample. This survey has already confirmed the planetary nature of Kepler-91b \citep[KOI-2133b,][]{lillo-box14,lillo-box14c}, Kepler-432b \citep[KOI-1299b,][]{ciceri14}, and Kepler-447b \citep[KOI-1800b,][]{lillo-box15b}; and has detected the secondary eclipse of WASP-10b \citep{cruz14}. In the present paper, we analyze 13 specific candidates in our sample that have shown either large RV variations or broadened spectra owing to the fast rotation of the host stars. Among the studied candidates, we have found an eclipsing binary that shows heartbeat-like modulations on its light curve. This effect was explained and explored by \cite{kumar95} and is due to tidal distortions in eccentric binary systems. The first object discovered whose light curve is mainly dominated by this type of modulation was KOI-54 \citep{welsh11}.  Since then, only a few more cases have been discovered \citep[e.g.,][]{thompson12}, but more cases are needed to understand and test theoretical models of how pulsations are induced from the dynamic tidal distortion. Interestingly, the RV analysis of KOI-3853 also shows indications of a possible third companion.

This paper is organized as follows. In Sect.~\S\ref{sec:observations}, we describe the observations used in this analysis. Section~\S\ref{sec:results} shows the analysis and results of the RV data for the found undiluted binaries (Sect.~\S\ref{sec:binaries}). Also, we set upper limit masses for the companions of few fast rotators (Sect.~\S\ref{sec:fastrotators}) and study two yet unsolved cases with current data in  Sect.~\S\ref{sec:unclear}. In Sect.~\S\ref{sec:lc} we show the detection of peculiar (not yet reported) features in the light-curves of some of the targets analyzed in this paper. Finally, a brief discussion and conclusions of these findings is provided in Sect.~\S\ref{sec:conclusions}.

%%%%%%%%%%%%%%%%%%%%%%%%%%%%%%%%%%%
%             		OBSERVATIONS
%%%%%%%%%%%%%%%%%%%%%%%%%%%%%%%%%%%

\section{Observations}

%--------------------------------------------------------
\subsection{{\it Kepler} photometry \label{sec:kepler}}

The objects analyzed in the present paper were firstly classified by the {\it Kepler} mission as potential planet hosts by the detection of planetary-like transits. We have used the long cadence data of quarters Q1-Q17 of the {\it Kepler} mission to analyze the light curves (LCs) of some of these {\it Kepler} objects of interest (hereafter KOIs). The data were obtained from the MAST\footnote{https://archive.stsci.edu/kepler/data\_search/search.php} (Milkuski Archive for Space Telescopes) database. We used the Pre-search Data Conditioning Simple Aperture Photometry flux \citep[PDCSAP][]{smith12, stumpe12} and detrend these data by using simple spline fitting to every set of data without jumps longer than one day. We have used one-day averaged flux bins as nodes for the spline fitting. In specified cases (such in KOI-3853), we removed some parts of the light-curve in the calculation of the nodes in order to not smooth or even eliminate non artificial modulations (such as primary transits, eclipses or other real variations).

%--------------------------------------------------------
\subsection{High-spatial resolution images\label{sec:astralux}}
{ The presence of undiluted sources can contaminate both the {\it Kepler} light curve and the spectra of the target, introducing imprints that could mask or mimic a planetary signal or contaminate the determination of its properties \citep[e.g. ][]{daemgen09, adams12, law13}. Thus,} prior to the RV observations, we obtained high-resolution images for the targets by using the lucky-imaging technique with the  AstraLux instrument at the 2.2m telescope in Calar Alto. In \cite{lillo-box12,lillo-box14} we already published the results of our high-spatial resolution survey of the {\it Kepler} candidates and provided details on the reduction and analysis of the data. More than 170 KOIs were observed and analyzed, including 9 out of the 13 KOIs studied here. For the remaining targets, we used ancillary high-resolution images from \cite{law13}  and the UKIRT J-band survey\footnote{http://keplergo.arc.nasa.gov/ToolsUKIRT.shtml} of the {\it Kepler} sample, publicly available in the CFOP and provided by D. Ciardi. 

In Table~\ref{tab:astralux} we summarize the main results obtained for each of the KOIs of the present study, including the calculated blended source confidence (BSC) and the number of detected visual companions closer than 6.0 arcsec. The BSC parameter was presented in \cite{lillo-box14b} and represents a standard and unbiased procedure to derive the probability of a given source to be isolated (i.e. without any blended undetected sources). { Additional sources were found closer then 6.0 arcseconds in just two KOIs (KOI-0131 and KOI-1463). However, these visual companions are too faint to affect the CAFE spectrum and thus their contribution to the RV is negligible. Added to this, they lie completely outside of the CAFE fiber (with 2.1 arcsec of diameter), thus not contaminating the spectra of the KOIs.}  

\begin{table}
\setlength{\extrarowheight}{2pt}
\small
\caption{Ancillary data and calculated BSC (blended source confidence) values for the KOIs studied in this paper.}             
\label{tab:astralux}      
\centering          
\begin{tabular}{r c c c c c}     % 7 columns 
\hline\hline       

 KOI      &     High-Res Im.\tablefootmark{a}  & Comp.  & Sep.     & $\Delta m$  & BSC          \\
          &     Ref.                           & ($<6$\arcsec)                   & arcsec.  & mag (band)  &  \%         \\ \hline 

\multicolumn{4}{l}{\bf Group A - Eclipsing binaries} \\ \hline
340.01   &     L13    & 0\tablefootmark{b} & &   & 99.99    \\
371.01   &     L13    & 0\tablefootmark{b} & &   & 96.0      \\     
686.01   &     LB14   & 0 & &                    & 99.9      \\     
3725.01  &     LB14   & 0 & &                    & 99.7        \\     
3853.01  &     LB14   & 0 & &                    & 99.6        \\     \hline
\multicolumn{4}{l}{\bf Group B - Fast rotators} \\ \hline
12.01    &     LB14   & 0 & &                    & 99.3        \\  
131.01   &     LB12   & 1 & 5.54\arcsec & $3.48$ ($i_{SDSS}$)    &  -     \\    
366.01   &     LB14   & 0 & &                 & 99.3      \\    
625.01   &     LB14   & 0 & &                    & 96.8      \\    
972.01   &     CFOP   & 0 & &                    &   -    \\    
%1020.01 &     L13    & 0 ($<2.5$\arcsec)   &   99.2    \\    
3728.01  &     LB14   & 0 & &                    & 99.6       \\      \hline    
\multicolumn{4}{l}{\bf Group C - Unsolved cases} \\ \hline
1463.01  &     CFOP   & 1 & 5.47\arcsec & $6.9$ (Kep.)   &  -    \\     
3890.01  &     LB14   & 0 & &                    & 98.3        \\     \hline
%2481.01  &     LB14   & 1 & $1.097$\arcsec  & $3.51$ ($i_{SDSS}$)  &   -   \\     
%$\pm 0.011$\arcsec
%\multicolumn{5}{l}{Fast rotators} \\
%\multicolumn{5}{l}{Unclear cases} \\

\hline                  
\end{tabular}
\tablefoot{Uncertainties in the angular separation of the companions and in their magnitude difference have been omitted for clarification purposes. We refer to the corresponding references for more information about this. 
\tablefoottext{a}{LB12 = \cite{lillo-box12}, L13 = \cite{law13}, LB14 = \cite{lillo-box14}, CFOP = no reference found in the bibliography but companion tables of the UKIRT J-band survey by D. Ciardi are available in the CFOP.}
\tablefoottext{b}{The maximum angular separation explored in \cite{law13} is 2.5 arcsec.}

}

\end{table}

%--------------------------------------------------------
\subsection{CAFE observations and data reduction \label{sec:observations}}

We have obtained multi-epoch RV data for the 13  KOIs shown in Table~\ref{tab:astralux}. We used the CAFE instrument \citep{aceituno13} mounted on the 2.2m telescope at Calar Alto Observatory (Almer\'ia, Spain) to obtain high-resolution spectra ($R=59000-67000$) in the optical range (4000-9000~\AA). Table~\ref{tab:observations} summarizes the targets observed, the number of epochs acquired, and the time span (defined as the time between first and last spectrum obtained) for each of them.

We have used the standard reduction process for the CAFE spectra, already described in \cite{lillo-box14c} and \cite{lillo-box14}. The basic reduction is performed by using the observatory pipeline described in \cite{aceituno13}\footnote{This pipeline was developed by Sebasti\'an S\'anchez.}. The wavelength-calibrated spectra are then analyzed with our own routines in order to obtain the RV.  In the case of the rapid rotators, we used the genetic algorithm {\it GAbox} \cite[presented in ][]{lillo-box14}\footnote{This algorithm has been already used to confirm the planetary nature of Kepler-91b in \cite{lillo-box14c,lillo-box14} and as a spectral energy distribution fitter in \cite{riviere-marichalar13,riviere-marichalar14}.} to determine the RV at the different epochs. We used the theoretical equivalency between maximum-likelihood and cross-correlation presented in \cite{zucker03} to match the observed spectrum to a synthetic template of the same spectral type. This RV extraction process is fully described in \cite{lillo-box14c}. 

For the slow rotators, a more precise analysis can be done by using the cross-correlation technique \citep{baranne96}. We used a binary weighted mask as template \citep{pepe02} and cross-correlate the observed spectra against it. The mask was designed by using a solar spectrum obtained from the solar atlas provided by BASS 2000\footnote{http://bass2000.obspm.fr/solar\_spect.php}. From this spectrum, we selected a series of isolated, sharp, and high-contrast absorption lines. In total, our mask includes around 2100 lines in the CAFE spectral range. We used a velocity range of $\pm30$ km/s around the expected RV value for each case. The RV of each spectrum is measured as the peak of a gaussian fit to the resulting cross-correlation function (CCF). This velocity is then corrected from the barycentric Earth radial velocity (BERV) obtained from the corresponding Julian date at mid-exposure.

In Tables~\ref{tab:KOI-0012} to \ref{tab:KOI-3890} of Appendix~\ref{app:A}, we provide the calculated radial velocities and the observing information (signal-to-noise, julian date, and phase) for each epoch and KOI.  

\begin{table*}[htbf]
\setlength{\extrarowheight}{1pt}
\small
\caption{Kepler Objects of Interest analyzed in this work. Observational data obtained  with CAFE at Calar Alto Observatory.}             
\label{tab:observations}      
\centering          
\begin{tabular}{r c c c c c c c c }     % 7 columns 
\hline\hline       

 KOI      & $m_{\rm Kep}$ & RA & DEC & Period\tablefootmark{(a)} & $M_{\star}$\tablefootmark{(b)} & $R_{\rm comp}$\tablefootmark{(c)} &Nspec & T$_{span}$         \\ 
          &   (mag.)       & (HH:MM:SS.SS) & (DD:MM:SS.SS) & (days) & ($M_{\odot}$) & ($R_{\odot}$)  & & (days)  \\  \hline
\multicolumn{8}{l}{\bf Group A - Eclipsing binaries} \\ \hline
\object{340.01} &  13.1  & 19:50:39.52    &     47:48:05.01   & $  23.673188\pm 0.000014  $  &    $1.10\pm0.07$   & $0.231\pm0.021^{\dagger}$  & 21 & 321.1  \\    
\object{371.01} &  12.2  & 19:58:42.28    &     40:51:23.37   & $ 498.3928\pm 0.0042      $  &    $1.55\pm 0.15$  & $0.77\pm0.39$  & 29 & 722.0  \\    
\object{686.01} &  13.6  & 19:47:21.78    &     43:38:49.64   & $  52.513565\pm 0.000015  $  &    $0.84\pm 0.12$  & $0.1248\pm0.0038^{\ddagger}$  & 19 & 39.0   \\   
\object{3725.01} &  10.1 & 19:40:48.69    &     38:31:10.38   & $ 1.5704970\pm 0.0000004\tablefootmark{(d)}  $  &    $2.52\pm 0.53$  & $0.825\pm0.007$  &  10 & 17.0 \\  
\object{3853.01} &  10.6 & 19:11:13.73    &     43:11:19.62   & $  21.512905\pm 0.000050  $  &    $1.55\pm 0.38$  & $0.31\pm0.22$  & 25 & 70.9  \\ \hline
\multicolumn{8}{l}{\bf Group B - Fast rotators} \\ \hline                                                              
\object{12.01}  &  11.4  & 19:49:48.90    &     41:00:39.56   & $  17.855149\pm 0.000010  $  &    $1.31\pm 0.27$  & $0.122\pm0.060$  & 47 & 114.0  \\   
\object{131.01} &  13.8  & 19:56:23.41    &     43:29:51.32   & $   5.014233\pm 0.000002  $  &    $1.31\pm 0.26$  & $0.082\pm0.029$  & 29 & 418.0  \\    
\object{366.01} &  11.7  & 19:26:39.40    &     38:37:09.32   & $  75.11202\pm 0.00010    $  &    $1.69\pm 0.17$  & $0.097\pm0.042$  & 44 & 72.9   \\   
\object{625.01} &  13.6  & 19:06:15.31    &     39:32:04.09   & $  38.13819\pm 0.00021    $  &    $1.46\pm 0.24$  & $0.43\pm0.21$  & 52 & 66.0   \\   
\object{972.01}  &  9.3  & 18:48:00.07    &     48:32:32.00   & $  13.118908\pm 0.000095  $  &    $2.02\pm0.07$   & $0.055\pm0.013$  & 51 & 335.3 \\  
\object{3728.01} &  12.3 & 19:35:05.31    &     44:38:18.49   & $   5.5460830\pm 0.0000013$  &    $2.05\pm 0.17$  & $0.1880\pm0.0009$  & 40 & 243.0\\  \hline
\multicolumn{8}{l}{\bf Group C - Unsolved cases} \\ \hline                                                           
\object{1463.01} &  12.3 & 19:13:02.07    &     43:22:35.27   & $ 1064.2681410\pm0.0001434$  &    $1.17\pm 0.28$  & $0.149 $  &  14 & 691.2 \\ 
%\object{2481.01} &  13.6 & 19:39:07.76    &     39:35:47.47   & $  33.84760\pm 0.00091    $  &    $1.62\pm 0.26$  &$0.\pm0.$  &  32 & 728.0 \\ 
\object{3890.01} &  13.2 & 19:09:52.29    &     37:57:59.90   & $ 152.82630\pm 0.00099    $  &    $1.76\pm 0.34$  & $0.3213\pm0.0091$  & 22 & 305.2 \\ \hline

\hline
\end{tabular}
\tablefoot{
\tablefoottext{a}{Transit periods are obtained from \cite{borucki11} and \cite{batalha13}.}
\tablefoottext{b}{Stellar masses are obtained from \cite{huber14}.}
\tablefoottext{c}{Radius of the transiting compnions according to \cite{batalha13}.$^{\dagger}$This radius was obtained from our own transit fitting (see Sect.~\ref{sec:lc0340}).$^{\ddagger}$In this case, the radius of the companion was obtained from \cite{diaz14}.}
\tablefoottext{d}{In Sect.~\S~\ref{sec:lc}, we have demonstrated that the actual period of this system is twice this value, i.e., $3.1409940\pm 0.0000008$ days.}

%\tablefoottext{a}{High-resolution image exisits. If so, the reference is provided: L12 = \cite{lillo-box12}, L13 = \cite{law13}, L14 = \cite{lillo-box14}, CFOP = no reference found in the bibliography but companion tables of the UKIRT survey by D. Ciardi are available in the CFOP.}
}

\end{table*}

%%%%%%%%%%%%%%%%%%%%%%%%%%%%%%%%%%%
%             		RADIAL VELOCITY
%%%%%%%%%%%%%%%%%%%%%%%%%%%%%%%%%%%

\section{Radial velocity analysis \label{sec:results}}

The extracted RV data explained in the previous section was analyzed by using the RVLIN software\footnote{http://exoplanets.org/code/} \citep{wright09} and its additional package BOOTTRAN for parameter uncertainties estimation with bootstrapping \citep{wang12} to fit our RV data to a Keplerian orbital solution. { The calculated uncertainties of the parameters account for both the RV and the stellar mass uncertainties.}{ In this section we study five eclipsing binaries (Group A), six fast rotating KOIs (Group B), and provide some hints for the presence of stellar-mass companions to two yet unsolved KOIs in our sample (Group C). { We note that the nature of the transit on KOIs of both Group B and C remains unsolved with the current data. In the former, due to the impossibility of getting accurate RV measurements and the later because of the lack of good phase coverage}. Tables \ref{tab:RVresults} and \ref{tab:FastRotators} summarize our RV analysis for the three groups of KOIs.} We also review the high-resolution images of these KOIs already published in \cite{lillo-box14b}.

%--------------------------------------------------------
\subsection{False positives: undiluted eclipsing binaries (Group A) \label{sec:binaries}}

Among the candidates being presented here, we have found five candidates that show large RV variations, incompatible with the presence of planetary-mass companions. These variations are modulated with the same period detected by the transit method using {\it Kepler} data. In the following, we analyze each of these false positives. In Table~\ref{tab:RVresults} we present the fitted orbital and physical parameters for those KOIs with a clear Keplerian fit to the RV. Figure~\ref{fig:rv} shows the measured radial velocities and the corresponding RV models of these KOIs\footnote{ KOI-3725 is not presented in this figure because its large $v\sin{i}$ prevents an accurate RV study.}.

\paragraph{KOI-0340.01 \vspace{0.3cm} \label{sec:rv0340} \\}

 This planetary candidate was announced by the {\it Kepler} team in the first release of the mission \citep{borucki10}, and still remains as a planetary candidate in the CFOP. Its {\it Kepler} light curve presents transit-like dims every $23.673188\pm 0.000014$ days. Our analysis of the {\it Kepler} light-curve of this KOI in Sect. \S~\ref{sec:lc} provides the detection of a secondary eclipse. Hence, we can keep the calculated period for the companion\footnote{Note that if differences in the odd/even transit depths are found we should double the period since we would be seeing the two eclipses of an eclipsing binary.}. 
 
This object was already included in the catalog of eclipsing binaries\footnote{http://archive.stsci.edu/kepler/eclipsing\_binaries.html} presented and updated in \cite{matijevic13}. Also, \cite{santerne12} used the SOPHIE instrument to obtain two RV measurements at quadrature phases  assuming circular orbit, and derived a semi-amplitude of $K=34.577 \pm 0.074$ km/s with such assumption, concluding that this system should be a single-line spectroscopic binary. Our CAFE data (obtained at eight different phases) provides a highly eccentric orbit ($e=0.513\pm0.005$) for the companion (in good agreement with the location of the secondary eclipse, see Sect.~\S~\ref{sec:lc0340}). The fitted RV semi-amplitude provides a value of $K=16.28\pm0.11$ km/s. Assuming a host mass of $M_{\star}=1.10\pm0.07\ M_{\odot}$ provided by the CFOP, we derived a stellar mass for the transiting object of $M_2\sin{i} = 0.214\pm0.006\ M_{\odot}$. The L-S periodogram indicates a period for the RV data in good agreement to the period obtained with the transits method thus indicating that both effects are produced by the same object (see Fig.~\ref{fig:rv}). Hence, the non-planetary nature of the object transiting KOI-0340 is clear from this analysis so that we can firmly and definitively discard this candidate as a planet. 

\paragraph{KOI-0371.01 \vspace{0.3cm}\\}
A periodic transit signal of $P_{\rm orb}=498.3928\pm 0.0042$ days was detected by {\it Kepler} for this star. No significant difference in the depth of the two single eclipses is found in this case, so we maintain the period of this KOI as calculated by the {\it Kepler} team. The analysis of our RV data (with a time span of 722 days) shows large relative variations for the stellar host. We find a clear (although broad) peak in the RV periodogram corresponding to the same periodicity as the transit signal (see Fig.~\ref{fig:rv}). According to our fit of the RV modulations, we find that the companion source should have a minimum mass of $M_2\sin{i}=0.445\pm0.014~M_{\odot}$. The physically bound stellar companion is calculated to be at $a=1.424\pm0.047$~AU, with a highly eccentric orbit of $e=0.407\pm0.008$. Thus, we can reject this planetary candidate and { establish} its binary nature.

\paragraph{KOI-0686.01 \vspace{0.3cm}\\}

This planet candidate was released in the second catalog of {\it Kepler} planet candidates. The RV analysis reveals large variations with a clear modulation. The analysis of the AstraLux images of this object (see Sect.~\S~\ref{sec:astralux}), yielded a 0.1\% of probability for this KOI having a non-detected blended background star. Thus, it is highly probable that these RV variations come from a bond companion to the KOI-0686 system.

The analysis of the detected transits by the {Kepler} team provided an orbital period of $P_{\rm orb}=52.513565\pm 0.000015$ days. 
If we assume the same period and mid-transit time for the RV signal, we can obtain a model solution for the RV data. This fit provides a stellar mass for the companion transiting object of $M_B\sin{i}=0.089\pm0.018~M_{\odot}$, thus confirming its stellar nature. We can thus confirm that KOI-0686 is an eclipsing binary. We note that a contemporary work by \cite{diaz14} also { obtained the same conclusion, with the fitted parameters  in good agreement within the uncertainties.}

\paragraph{KOI-3725.01 \vspace{0.3cm}\\}

The observations for this object were started while its status was "not-dispositioned" in the {\it Kepler} catalog. This meant that the light-curve analysis had not yet passed all the required criteria to be considered as a planet candidate. After the target was observed with CAFE, the {\it Kepler} team updated its status to "false positive". Our analysis of the light-curve in Sect. \S~\ref{sec:lc} shows a clear difference in the depth of the odd/even eclipses, thus tagging this system as an eclipsing binary. Given this, we must consider a double period than the published by the {\it Kepler} team { (i.e., $P_{\rm orb}=3.1409940\pm 0.0000008$ days instead of $1.5704970\pm 0.0000004$ days)}.

Our high-spectral resolution data (and light-curve analysis in Sect.~\ref{sec:lc}) confirms that this system is an eclipsing binary, showing large RV variations over the observed time span. However, the small number of phases acquired and the contaminated spectrum by the presence of two possible sets of lines, being one very broad ({ suggesting} large rotational velocities), prevents us from performing any detailed RV analysis.

\paragraph{KOI-3853.01 \vspace{0.3cm}\\}

This KOI { has been reported as a possible false positive in the Kepler database\footnote{http://exoplanetarchive.ipac.caltech.edu}}. The derived period according to the light curve analysis is $P_{\rm orb} = 21.512905\pm 0.000050$ days. The analysis of our high-resolution spectra obtained in 14 different phases reveals a large RV variation. When fitting the RV with a single object, an important periodic trend is found in the residuals. We tried to fit the RV data by assuming the presence of a third body in the system (see Fig.~\ref{fig:koi3853}). The fit in this case was significantly improved. We can measure the significance of such improvement by using the Bayesian Inference Criterion\footnote{$\Delta BIC=(\chi_1-\chi_2)-(k_1-k_2)\ln N$, where where $\chi^2_1$ and $k_1$ are the $\chi^2$ statistic and the number of free parameters for the one-companion model and $\chi^2_2$ and $k_2$ are the $\chi^2$ statistic and the number of free parameters for the two-companions model. $N$ is the number of RV points.} (BIC). We obtain a difference in the BIC value of $\Delta$BIC $>100$, showing a very strong evidence for the presence of a third companion. 

Interestingly, the period of the less massive companion (i.e., component C) is similar (in agreement within $3\sigma$ uncertainties) to the more massive companion (i.e., component B), thus being in a near 1:1 mean motion resonance (MMR). {  The periodogram of the RV data shows a double (blended) peak around the period corresponding to the the eclipse observed in the {\it Kepler} data. Both peaks have a false alarm probability (FAP) below 1\%. In order to test the validity of the second periodicity in the data, we fixed the fitted parameters obtained for component B and we ran $10\times 10^5$ MCMC chains to fit the residuals of the data. We let the period of the possible C component to vary in the range 1-71 days. The posterior distribution of the combined chains clusters around the estimated period provided by the joint fit (within the uncertainies). }

{ Thus all these three tests performed to our RV data (BIC, periodogram, and MCMC analysis) favor the existence of a third component in 1:1 MMR against the two-bodies only configuration.} The orbits of the two possible companions would be oriented in opposition, i.e., the difference in the argument of the periastron of both eccentric orbits agrees with being 180$^{\circ}$. According to the fitted models, and assuming the stellar mass for the A component provided in Table~\ref{tab:observations}, the minimum masses of the two companions would be $M_B\sin{i_B} = 0.527\pm0.044~M_{\odot}$ and $M_C\sin{i_C} = 0.030\pm0.016~M_{\odot}$. Note that the mass difference between both components is significantly large, with the less mass component in the substellar regime.

The unusual architecture of the proposed scenario would require a more detailed study of its stability (which is beyond the scope of the present paper). Simulations of 1:1 mean motion resonances in the planetary-mass regime by \cite{antoniadou14} showed that these configurations are stable under specific mass ratio conditions, $m_C/m_B<0.0205$. Although our objects are more massive, we note that this condition is fulfilled (within the uncertainties) for the KOI-3853 system. However, as stated before, more work is needed to analyze the stability of this system. 

{ Stellar pulsations induced by the companion could also explain the correlated residuals in the RV fit. As we will see in Sect.~\ref{sec:lc}, the light-curve of this target presents the so-called {\it heartbeat} effect, produced by tidal interactions induced by the more massive stellar companion. In \cite{willems02}, the authors describe and calculate the RV variations induced by tides in close binary systems. According to them, there  are specific resonant configurations that can lead to radial velocity variations even larger than 5 km/s, thus being a relevant explanation for the presence of the modulated residual of this close binary. However, an accurate description of this system is out of the scope of this work. Future analysis studying the heartbeat-like modulation and the stellar modes induced by the close massive companion could unveil or reject the existence of the substellar companion in the 1:1 resonance orbit.}

\begin{table*}
\setlength{\extrarowheight}{2pt}
\scriptsize
\caption{Results of the fitting to the observed radial velocities.}             
\label{tab:RVresults}      
\centering          
\begin{tabular}{r c  c c c c c c c c}     % 7 columns 
\hline\hline       

KOI                &     K          &  e  & $\omega$   & V$_{sys}$   & $M_{\rm comp}\sin{i}$\tablefoottext{a} & inc.\tablefoottext{b} & $F_M$\tablefoottext{c} & a  & F-test\tablefoottext{d} \\
                   &   (km/s)       &     &   (deg.)   & (km/s)   & ($M_{\odot}$)  & (deg.)&  ($M_{\rm Jup}$) & (AU) &  \\ \hline

\multicolumn{8}{l}{\bf Group A - Eclipsing binaries} \\ \hline
      0340 B  &       16.28$\pm$0.11  &   0.5138$\pm$0.0059  &     239.71$\pm$0.58  &    -84.007$\pm$0.045  &   0.2144$\pm$0.0058  &  $89.95$  &   7.000$\pm$0.091  &   0.1614$\pm$0.0054 & $>100$ \\
      0371 B  &      8.252$\pm$0.086  &   0.4070$\pm$0.0081  &      94.75$\pm$0.47  &    -56.855$\pm$0.081  &     0.445$\pm$0.014  &  $89.95$  &    23.15$\pm$0.82  &     1.424$\pm$0.047 & $>100$ \\
      0686 B  &          6.3$\pm$1.7  &       0.52$\pm$0.15  &           63$\pm$12  &      -31.30$\pm$0.65  &     0.089$\pm$0.018  &  $89.38$  &     0.88$\pm$0.45  &     0.259$\pm$0.015 & $>100$ \\
      3853 B  &       26.98$\pm$0.48  &   0.3956$\pm$0.0084  &       130.8$\pm$1.3  &      -75.19$\pm$0.15  &     0.527$\pm$0.043  &  $50.42$  &      35.5$\pm$2.2  &     0.175$\pm$0.014 & $>100$ \\ 
      3853 C  &       2.1$\pm$1.3     &   0.56$\pm$0.20      &       312$\pm$21     &      -75.19$\pm$0.15  &     0.030$\pm$0.016  &  $-$  &      0.011$\pm$0.019  &     0.171$\pm$0.014 & $>100$ \\ \hline
\multicolumn{8}{l}{\bf Group C - Unsolved cases} \\ \hline
       1463 B  &           $>3.3$  &          $>0.18$  &     [80,200]      &        [-28.5,-25.5]  &       $>0.183\pm0.030$  &   $89.95$  &   $>4.15$  &     $2.15 \pm 0.19$ & $>100$ \\
%\tablefoottext{c}\       2481 B  &       190.7$\pm$2.5  &       0.00  &         -  &         5.1$\pm$1.1  &      15.65$\pm$0.46  &     0.204$\pm$0.050  &         2.6$\pm$1.3  &     0.762$\pm$0.042 \\
%\tablefoottext{c}\       2481 Ab &      33.85$\pm$0.43  &       0.00  &         -  &       0.48$\pm$0.13  &      15.65$\pm$0.46  &   0.0101$\pm$0.0028  & 4.0$\pm$2.4)$\times 10^{-4}$  &     0.241$\pm$0.013 \\
       3890 B  &        $>2.5$  &    $>0.33$  &    [0,82]  &      [-28.7,-24.2]  &       $>0.097\pm0.014$  &   $89.95$  &   $>0.27$  &    $0.676\pm0.040$    & 3.6 \\

\hline                  
\end{tabular}
\tablefoot{
\tablefoottext{a}{Companion mass assuming primary mass from \cite{huber14}, see Table~\ref{tab:observations}.}
\tablefoottext{b}{Orbital inclination provided by the {\it Kepler} team.}
\tablefoottext{c}{Mass function, defined as: $F_M=M_{\rm comp}^3\sin^3{i}/(M_{\rm comp}+M_{\star})^{2}$. This parameter is directly obatined from the RV fit and does not depend on the primary mass.}
\tablefoottext{d}{F-test comparing the significance of the RV model to a straight line.}
%\tablefoottext{d}{The period of this object corresponds to component A2 (see Sect.~\ref{sec:unclear}). The period of the transiting object is $33.84760 \pm 0.00091$ days.}
}
\end{table*}

%  23.673$\pm$0.012  &
%     498.4$\pm$3.5  &
%      52.5$\pm$2.5  &
%  21.512$\pm$0.041  &
%  20.75$\pm$0.26    &

%áááááááááááááááááááááááááááááááááááááááááááááááááááááááá
% 			Subsection 
%áááááááááááááááááááááááááááááááááááááááááááááááááááááááá
\subsection{Fast rotators (Group B). Upper limits to companion masses  \label{sec:fastrotators}}

Planets orbiting fast rotating stars are more difficult to detect by the RV method. The line broadening produced by this effect prevents an accurate measurement of the line shift, increasing significantly the uncertainties for large $v\sin{(i)}$ values. Six fast rotators ($v\sin{(i)}>20$ km/s) were found in our sample. We computed  individual spectral templates with the same physical properties (effective temperature, surface gravity, metallicity, and rotational velocity) as the KOI by using the ATLAS09\footnote{http://kurucz.harvard.edu/grids.html}. Then {\it GAbox} was used to compare the template and the observed spectra and determine the shift between both (i.e., the RV), by following the theoretical equivalency between maximum-likelihood and cross-correlation presented in \cite{zucker03}\footnote{See \cite{lillo-box14c} for a complete description of the method with {\it GAbox}.}.

As stated, we can just provide an upper limit to the mass of the companion of these KOIs. We have assumed circular orbits for all of them{ , as well as the lack of blended companions contaminating the RV data (see Table~\ref{tab:astralux})}. { A simple circular model was fitted to the RV data by fixing the period and time of mid-transit calculated by the {\it Kepler} team from the analysis of the eclipse. Thus, only two free parameters were fitted, the RV semi-amplitude ($K$)  and the systemic velocity ($V_{\rm sys}$). We ran one MCMC chain of $10^6$ steps and discarded the first 10\% to avoid dependency with the priors. The results for all six fast rotators are presented in Table~\ref{tab:FastRotators}. According to these results, all KOIs but KOI-3728 are compatible with having $K=0$~km/s (non detection) within the uncertainties. Thus, we can take $\sigma_{RV}=K+3\times\sigma_K$ (where $\sigma_K$ is the uncertainty in $K$) as the upper limit of the semi-amplitude of the RV variations caused by a non-detected companion. In the case of KOI-3728, the solution for $K$ is not compatible with zero, thus suggesting a possible detection of the RV variations provoked by the transiting companion. According to this analysis, the companion would have a mass of  $0.068\pm0.041~M_{\odot}$. However, due to the large uncertainties and for consistency with the structure of this paper, we prefer to maintain this KOI in this group B, although highlighting the possible detection of a stellar-like companion. The large $v\sin{i}$, instead, does not allow us to ensure such detection.

Hence, the RV amplitude ($K$) induced by the companion (if any) detected in these KOIs should be smaller than the aforementioned upper limit ($\sigma_{RV}$), i.e. $K<\sigma_{RV}$. This amplitude is defined for circular orbits as 

\begin{equation}
K^3 = \frac{2\pi G}{P} \left(\frac{M^3_{\rm comp} \sin^3(i)}{(M_{\rm comp}+M_{\star})^2}\right),
\end{equation}

\noindent where the term in parenthesis is called the mass function ($F_M$), $M_{\star}$ is the stellar mass, $M_{\rm comp}$ is the mass of the companion, and $P$ its orbital period.} We can now determine the maximum mass of the companion that accomplishes $K=\sigma_{RV}$. This value represents the maximum mass ($M_{\rm comp}^{max}\sin{i}$) of the hypothetic companion orbiting the star and inducing the detected transit signal. The calculated maximum masses  of the hypothetic companions ($M_{\rm comp}^{max}\sin{i}$) are listed in Table~\ref{tab:FastRotators}. 

{ All but two KOIs (KOI-0972 and KOI-3728) have upper mass limits below the substellar regime ($\sim 80~M_{\rm Jup}$). Indeed, in one case (namely KOI-0366), the upper mass limit is below the deuterium burning limit ($\sim 13~M_{\rm Jup}$), suggesting that the object transiting this star is likely a planet.}

\begin{table}
\setlength{\extrarowheight}{5pt}
\scriptsize
\caption{Upper limits for the mass of the transiting companion to KOIs with large $v\sin{i}$ values (fast rotators, Group B).}             
\label{tab:FastRotators}      
\centering          
\begin{tabular}{r c c c c c}     % 7 columns 
\hline\hline       

%\multicolumn{4}{l}{\bf Group B - Fast rotators} \\ \hline
%KOI &  $v\sin{i}$   &$\sigma_{RV} $   & $ M_{\rm comp}^{max}\sin{i}$ \\ 
%     12.01     &      $70\pm5$\tablefootmark{a}        &    &   (km/s)      &  (km/s)         &     ($M_{\rm Jup}$)          \\ \hline
%    131.01     &      26$^{\dagger,}$\tablefootmark{b}   2.02  $\pm$    0.36       &   31.6   $\pm$    6.9  \\
%    366.01     &      $35\pm2$\tablefootmark{c}        &      &  4.2  $\pm$     1.3       &     43   $\pm$     14  \\
%    625.01     &      $25\pm2$\tablefootmark{c}         0.768  $\pm$  0.077       &   22.8   $\pm$    2.6  \\
%    972.01     &      $120\pm5$\tablefootmark{c}        & 1.06  $\pm$    0.21       &   22.7   $\pm$    5.3  \\
%   3728.01     &      55$^{\dagger,}$\tablefootmark{a}  & 4.63  $\pm$    0.85      &     88   $\pm$     17  \\
%      &  9.3  $\pm$     1.9       &    136   $\pm$     29  \\ 
%   \tablefootmark{a}2481.01     &    4         & 0.147 $\pm$ 0.018   &   3.23  $\pm$ 0.5 \\
 %  1020.01    &      16       &   60  $\pm$      47       &   2200   $\pm$   2000  \\

\multicolumn{4}{l}{\bf Group B - Fast rotators} \\ \hline
    KOI     &                         $v\sin{i}$   &  $K$                     &    $V_{sys}$        & $\sigma_K$  &   $ M_{\rm comp}^{max}\sin{i}$  \\ 
            &                                      &   (km/s)                 &  (km/s)             &  (km/s)    &   ($M_{\rm Jup}$)          \\ \hline
  12.01     &      $70\pm5$\tablefootmark{a}        &       $0.140\pm1.425$   &    $-18.6\pm1.1$  &      1.565   &   $ 25.2\pm 3.7$  \\
 131.01     &      26$^{\dagger,}$\tablefootmark{b}  &       $1.400\pm1.995$   &    $ -9.0\pm1.8$  &      3.395   &  $ 35.8\pm 5.0$  \\
 366.01     &      $35\pm2$\tablefootmark{c}        &       $0.027\pm0.255$   &    $  8.5\pm0.2$  &      0.282   &   $ 8.70\pm0.59$  \\
 625.01     &      $25\pm2$\tablefootmark{c}        &        $0.120\pm0.510$   &    $-25.87\pm0.42$  &      0.630   &  $ 14.1\pm 1.6$  \\
 972.01     &      $120\pm5$\tablefootmark{c}       &        $2.200\pm3.300$   &    $-14.1\pm2.4$  &      5.500   &  $106.8\pm 2.6$  \\
3728.01     &      55$^{\dagger,}$\tablefootmark{a}  &       $4.810\pm2.775$   &    $-52.6\pm2.4$  &      7.585   &  $111.6\pm 6.3$  \\

\hline                  
\end{tabular}
\tablefoot{$^{\dagger}$ No uncertainties provided.
\tablefoottext{a}{Values obtained from the analysis of TrES spectra by A. Byerla.}
\tablefoottext{b}{Values obtained from the analysis of Lick spectra by H. Isaacson.}
\tablefootmark{c}{Values obtained from the analysis of McDonald or TrES spectra by S. Quinn.}
}
\end{table}

%áááááááááááááááááááááááááááááááááááááááááááááááááááááááá
% 			Subsection 
%áááááááááááááááááááááááááááááááááááááááááááááááááááááááá
\subsection{Unsolved cases (probable false positives,  Group C) \label{sec:unclear}}

\paragraph{KOI-1463.01\vspace{0.3cm}\\}

{ Only four phases (14 spectra in total) have been obtained with CAFE for this long-period candidate ($P_{\rm orb}=1064.2681410
\pm0.0001434$ days). }

Our RV data (with a timespan of 691 days) shows large variations of $\sim 7.5~$km/s, not corresponding to a planetary-mass object (see Fig.~\ref{fig:rv2}, left panel). However, the low number of epochs acquired and the timespan (shorter than the orbital period) prevents us from extracting definite conclusions about this system. { We have used a radial velocity model including non-zero eccentricity to fit the available data. The free parameters used were thus the RV semi-amplitude, the eccentricity, the argument of the periastron, and the systemic velocity of the system. 

We ran a Monte-Carlo Markov-Chain with $10^6$ steps to estimate these parameters and detect the correlations between them due to the lack of a complete sampling of the orbital phases. The degeneracy in the parameter space is important but we must remark that all solutions provide eccentricities larger than $e>0.18$ and semi-amplitude values larger than $K>3.3$~km/s. This corresponds to a lower-mass limit for the companion transiting object of $M_B^{min}=0.183\pm0.030~M_{\odot}$. In the left panel of Fig.~\ref{fig:rv2}, we show a random sample of solutions obtained for this system. 

Thus, the transiting companion, according to the present data is likely of stellar nature, suggesting that KOI-1463.01 is a false positive.} The solution to this system is still open and more RV data in a larger time span (since {\it Kepler} will no longer observe this target) will help solving this dichotomy, unveiling the nature and properties of the transiting object. We also note that \cite{hirano12} suggested that this candidate could actually be a late-type stellar companion transiting the star due to their analysis of the flux variations.

\paragraph{KOI-3890.01 \vspace{0.3cm}\\}

This planet candidate was detected to transit its host star with a periodicity of $152.82630\pm0.00099$ days. Our RV analysis, in a time span of 305.2 days shows a clear trend with a variation of 3 km/s (see Fig.~\ref{fig:rv2}, right panel). Since we do not detect transit depth differences in the odd/even transits, we can fix the period and mid-transit time to that measured in the light-curve analysis by the {\it Kepler} team. { Also, due to the small range of phases covered by our observations ($\phi \in [0.6,0.8]$), we cannot constrain the eccentricity and argument of the periastron of the orbit. We then proceed in the same manner as with KOI-1463, running an MCMC chain and analyzing the range of parameters. By doing so, we obtain a minimum RV semi-amplitude of $K>2.5$~km/s. This translates into a minimum mass for the transiting companion such that $M_B>0.097\pm0.014~M_{\odot}$, with a minimum eccentricity of $e>0.33$. In the right panel of Fig.~\ref{fig:rv2}, we show a random sample of solutions for this degenerated system. 

However, we must  warn that in this case, the F-test comparing the RV model to a simple straight line  provides a value of 3.6, slightly favoring the false positive detection against a null detection model. Thus, although this candidate is likely a binary, we cannot extract definite conclusions given the current data (mainly due to the small phase coverage).  More RV at different phases is necessary to better understand this system.}

%%%%%%%%%%%%%%%%%%%%%%%%%%%%%%%%%%%
%             		LIGHT CURVE ANALYSIS
%%%%%%%%%%%%%%%%%%%%%%%%%%%%%%%%%%%

\section{Light-curve analysis \label{sec:lc}}

In this section we analyze the {\it Kepler} light-curves of the targets studied in this work to look for possible interesting features that can contribute to better characterize these systems (secondary eclipses, photometric natural modulations - ellipsoidal, beaming, or reflexion-, and signatures of possible additional objects) in the manner as we did in Kepler-91 \citep{lillo-box14}. { We note that the accuracy of some parameters (such as the eccentricity) could be benefited from a  joint analysis of the RV and the LC. However, in the particular cases studied in this paper, the eccentricity of the orbits are well constrained with the RV data (since most of the orbits are well sampled) and we assume the period and time of transit from the analysis of the LC by the Kepler team. Thus, we consider that the joint analysis would not improve the results significantly in the present cases.}

In the next subsections, we individually analyze the results of the targets were additional effects (rather than the detected transit by {\it Kepler}) have been found.

%+++++++++
\paragraph{KOI-0340 \vspace{0.3cm}\\ \label{sec:lc0340}}

We have detected the secondary eclipse (see Fig.~\ref{fig:lc0340}) of this RV-identified binary orbiting in a highly eccentric orbit (see Sect. \S~\ref{sec:rv0340}). We have used a simple fitting model of a binary box { with linear ingress and egress dependencies} (enough for the purposes of the paper) to measure its depth, location, and duration. { The period and time of mid-eclipse was assumed from that calculated by the {\it Kepler} team.} The results show a depth of $\delta_2=650.6\pm6.6$~ppm, located at $\phi=0.31862\pm0.00002$, and lasting $d = 0.01005\pm 0.00023$ in phase units, equivalent to $d = 5.70\pm 0.13$ hours assuming the period provided by {\it Kepler} team. By assuming the primary eclipse depth provided by the {\it Kepler} team ($\delta_1 = 22335$~ppm, no uncertainty provided), we can derive a surface temperature ratio with 
\begin{equation}
\label{eq:TempRatio}
\frac{\delta_1}{\delta_2}=\left(\frac{T_A}{T_B}\right)^4,
\end{equation}
\noindent where $\delta_1$ is the depth of the deepest eclipse, $\delta_2$ is the depth of the smaller eclipse, $T_A/T_B$ is the ratio of effective temperatures of both stars. By using this equation, we obtain a temperature ratio of $T_A/T_B=2.421\pm0.006$.

Also, as expected due to the large eccentricity and orientation of the orbit (obtained by the RV analysis), the location of the secondary eclipse is not centered at mid-time phase but instead at $\phi=0.31862\pm0.00002$. The expected phase difference is known to be given by \citep{wallenquist50}

\begin{equation}
\Delta\phi = 0.5+e\cos{\omega}\frac{1+(1/\sin{i})^2}{\pi}
\end{equation}

According to the parameters obtained with the RV analysis we would have expected the secondary eclipse to be located at $\phi = 0.335\pm0.010$,  if we assume the inclination provided by the {\it Kepler} team. The difference between the expected time of mid-eclipse and the measured time is $9.3\pm5.7$ hours. { The source of this discrepancy is unknown but could be due to an underestimation of the uncertainties in the RV parameters or an inaccurate determination of the time of eclipse, translated into a small shift in the phase-folded light curve.}

{ Due to the large eccentricity found for this system in Sect.~\ref{sec:binaries}, we decided to perform a dedicated modeling of the primary eclipse\footnote{The {\it Kepler} team assumes circular orbits in the transit fitting.}.  We used the \cite{mandel02} light-curve models and obtained the quadratic limb-darkening parameters by interpolating the stellar parameters provided by \cite{huber14} to the values in \cite{claret11} for the {\it Kepler} band. We fixed the eccentricity and argument of the periastron to the values found in the radial velocity analysis and leave the inclination ($i$), semi-major axis to primary radius ($a/R_{A}$) and radius ratio ($R_A/R_B$) as free parameters. The results for this fit are: $i = 89.6^{\circ}\pm0.2^{\circ}$, $a/R_A=21.82\pm0.34$, and $R_B/R_A= 0.1452\pm0.0007$. This model is shown in the upper panel of Fig.~\ref{fig:lc0340}. { The main difference with the {\it Kepler} team parameters ($i_{\rm Kep} = 89.95^{\circ}$, $(a/R_A)_{\rm Kep}=14.31983\pm0.00722$, and $(R_B/R_A)_{\rm Kep}= 0.142771^{+0.000068}_{-0.000048}$) is the semi-major to stellar radius ratio. This is due to their assumption of circular orbit, while we have found a relatively large value for the eccentricity from both the RV and the secondary eclipse analysis. }

%+++++++++
\paragraph{KOI-3725 \vspace{0.3cm}\\}

The most remarkable imprint in the light-curve of KOI-3725 when phase-folding it with the period provided by the CFOP is the clear presence of two sets of eclipses with different depths. If we phase-fold the LC with twice the current period, we obtain a deeper eclipse at phase $\phi=0$ than at $\phi=0.5$. This is a clear sign of an eclipsing binary. { According to this, the actual periodicity of this binary system is $P_{\rm orb}=3.1409940\pm 0.0000008$.}

The temperature ratio between both stars can be estimated from the depth ratio of both eclipses following Eq.~\ref{eq:TempRatio}. In this case, a simple fitting to the gaussian-like eclipses (enough for the purposes of this paper) provides $\delta_1=2732\pm30$~ppm and  $\delta_2=1803\pm35$~ppm (see fitted functions over plotted in the bottom panel of Fig.~\ref{fig:lc3725}). This implies a temperature ratio of $T_A/T_B=1.109\pm0.006$.

%+++++++++
\paragraph{KOI-3853 \vspace{0.3cm} \label{sec:lc3853}\\}

The light-curve of KOI-3853 shows an interesting effect just discovered in a few number of tidally distorted binaries, the {\it heartbeat} pulsation.  The shape of the light-curve modulation due to this effect is well understood and was parametrized by \cite{kumar95}. In particular, in the case of KOI-3853, its shape corresponds to a highly inclined orbit with the periastron oriented  behind the plane of the sky. An accurate model fitting of this modulation is beyond the scope of this paper. Future works will focus on this object and fully characterize it.

However, the {\it heartbeat} is not the only imprint in the light-curve of this system. A transit-like shape appears in the middle of the tidal modulation in the phase-folded light-curve. The {\it Kepler} team identified this transit as a potential candidate being of planetary nature. In the lower panel of Fig.~\ref{fig:lc3853}, we show a zoom to this region once the "heartbeat" effect has been removed by a simple cubic spline fitting. The eclipse has a maximum depth of around 2000 ppm. However, it is filled by some data points at lower depths.  The period of both the heartbeat signal and the eclipse { are exactly the same, indicating that both effects are provoked by the same object (i.e., companion B)}. Interestingly, we can see both a duration and depth variation in the individual eclipses (see Fig.~\ref{fig:lc3853EclipseDepths}). Figure~\ref{fig:lc3853eclipses} displays the individual (detrended) eclipses. 

Our RV analysis (see Sect.~\ref{sec:binaries}) { suggests that} the motion variations of the primary star (A) are well described with the presence of two companions, a massive stellar object (component B) and a sub-stellar companion (component C) in a near 1:1 mean motion resonance. Both objects would revolve in eccentric orbits around the primary (massive) star. Based on this, we here suggest some possible explanations for the depth and duration variation of the detected eclipses:

\begin{itemize}
\item \emph{Orbital wobble due to satellite companion}. The presence of a third body orbiting component B can periodically vary its impact parameter, thus changing both the duration and the depth of the eclipse. Since we know from the shape of the heartbeat effect that the orbital inclination should be high, this should be a grazing eclipse of component B over component A. In the first periods of Fig.~\ref{fig:lc3853eclipses}, the projected location of component B in the plane of the sky would be at impact parameters $b\gtrsim 1+R_B/R_A$ so that we see a small or no eclipse. By contrast, in the last periods observed, the impact parameter decreases so that the eclipse is deeper. A longer time span could have confirmed this hypothesis by detecting periodic variations of the eclipse depth.

\item \emph{Orbital wobble due to component C}. In our RV analysis we found that the observed RV signal is much better modeled with two co-orbital stars with the same periodicity (in 1:1 mean motion resonance) but in two eccentric orbits in opposition, i.e., with their pericenters differing by 180$^{\circ}$. Component C, maybe in a higher inclined orbit (since we do not see any other dip in the LC), could be perturbing the orbit of component B, provoking an orbital wobble able to change its impact parameter and thus the eclipse depth.

\item \emph{Orbital circularization}. If the system is being circularized, with component B orbiting closer to component A every period, the eclipse depth and duration will be increased with time. Since within the time span of {\it Kepler} data we do not see a new turn back to smaller depths, we cannot discard this possibility.

\end{itemize}

With the current data, none of the previous conclusions can be discarded so that the problem remains open. However, our RV analysis suggests the presence of a third component in near 1:1 mean motion resonance. Future observations and study will bring more light and try to unveil the nature of this system. In particular, a dedicated stability study is required to check the dynamical plausibility of the co-orbiting substellar object in 1:1 mean motion resonance. As we mentioned in Sect.~\S~\ref{sec:binaries}, the mass ratio between the two companions would be in agreement to the allowed ratio for a triple system in 1:1 mean motion resonance as calculated by \cite{antoniadou14}, although we warn that this study was specifically performed for planetary-mass systems.

%%%%%%%%%%%%%%%%%%%%%%%%%%%%%%%%%%%
%             		CONCLUSIONS
%%%%%%%%%%%%%%%%%%%%%%%%%%%%%%%%%%%

\section{Summary and discussion \label{sec:conclusions}}

{ We have analyzed the RV data obtained with the CAFE spectrograph for 13 KOIs and the {\it Kepler} light-curves of some of them. In the majority of the eclipsing binaries found in this work (Group A), we found that the lighter component is a very low-mass stars (VLMSs). \cite{zhou14} and \cite{diaz14} published the last results on this kind of objects (see also references there in), adding more examples to the reduced number of well-characterized VLMSs (with detected eclipses and RV variations). Added to this, our RV analysis suggests substellar masses for the transiting companions of the fast rotating stars (Group B). 

In Fig.~\ref{fig:MR}, we show the mass-radius diagram for the companions found around these KOIs. In the case of KOI-0340 B, the radius provided by the {\it Kepler} team was well below the expected value provided by any isochrone. We then decided to make a dedicated primary eclipse fit (see Sect.~\ref{sec:lc0340}) accounting for the large eccentricity measured in Sect.~\ref{sec:binaries}. By assuming the primary radius obtained by \cite{huber14}, the new radius of the companion, although larger, still lies below the expected value. This could be attributed to an underestimation of the radius of the primary star ($R_A$). We can estimate $R_A$ from the $a/R_A$ ratio obtained by the primary eclipse fitting and the $a$ value obtained in the RV analysis. The resulting updated radius is $R_A = 1.589\pm0.021~R_{\odot}$. If we now use this value and the radii ratio between both components obtained from the eclipse fitting ($R_B/R_A$), we obtain $R_B=0.231\pm0.021~R_{\odot}$. This new value perfectly agrees with the expected value for the calculated mass of the companion. 

In the case of the fast rotator KOI-0972, the calculated upper mass limit for its transiting objects (assigning this mass limit to every candidate) also lie below the isochrones of VLMSs. The {\it Kepler} team detected two planet candidates in this system, with measured radii $R_{.01} = 6.1\pm1.4~R_{\oplus}$ and $R_{.02} = 1.73\pm0.57~R_{\oplus}$. There are thus two explanations for the discrepancy in the location of these objects in the mass-radius diagram: i) an underestimation of the stellar radius, or ii) a real planetary nature for the candidates, with corresponding masses well below our upper limits. The former explanation is plausible since the host seems to be a sub giant star and the Darmouth models used to determine the radius could be biased for this kind of stars \citep[see ][]{huber14}. The latter possibility is represented in Fig.~\ref{fig:MR} by the shadowed regions. We can see that they would agree with having planetary-like densities, in agreement with their measured radii, if their masses are well below our detection limits.

{According to the mass-radius diagram and isochrones, we can also obtain an estimation of the radius of the suggested third component in KOI-3853 (i.e., KOI-3853 C). By assuming the calculated mass obtained by RV and the isochrones between 1-5 Gyr, the radius of this object should be in the range $R_C \approx 0.08-0.11~R_{\odot}$. } 

In the following sections we summarize the results of each group of KOIs:}

\subsection{Eclipsing binaries (False positives, Group A)}
We have spectroscopically identified five eclipsing binaries in the sample of {\it Kepler} planet candidates via RV analysis. Among them, KOI-0340 was already identified as a binary and KOI-3725 was initially classified as "not-dispositioned" but then updated to "false positive" after we had already obtained the spectra. We detected the secondary eclipse of KOI-0340 using {\it Kepler} data. By measuring its depth, we concluded that the temperature ratio of both stars is $T_{\rm eff}^A/T_{\rm eff}^B=2.45\pm0.03$. Also, the location of the secondary eclipse agrees (except for a small offset) with the high eccentricity of the orbit derived by the RV analysis. In the case of KOI-3725, we concluded that the period of the system is actually twice that published by the {\it Kepler} team, since we detect clear differences in the odd/even eclipse depths. The location of the secondary (less deep) eclipse confirms a circular orbit for this system. 

 { Apart from the already confirmed false positive KOI-0686, we have reported other two eclipsing binary systems in this work (namely KOI-0371.01 and KOI3853.01). They present RV variations with the same periodicity as the transit signal.} The case of KOI-3853 is of particular interest. Its light-curve shows the clear signs of a recently discovered class of tidally distorted binaries, also known as \emph{heratbeat} stars. We have found that the eclipse detected by {\it Kepler} changes progressively in depth and duration. The high inclination of the orbit of the system (known because of the shape of the heartbeat effect) agrees with this dim being a grazing eclipse of the two stars. Our RV data suggests the presence of a third (substellar) component in the system revolving in an eccentric orbit around the primary star in a near 1:1 mean motion resonance with the more massive detected companion. Taking this into account, we hypothesize three possible explanations for the depth and duration variations: i) a third body perturbing the impact parameter on every transit, ii) an orbital wobble of component B due to the perturbing substellar companion, and iii) a possible circularization of the orbit with the stars getting closer and so decreasing the impact parameter. This dichotomy could be solved with a larger photometric time span. In the first and second scenarios we should see a modulation of the eclipse depth while in the second we should just see a rise in the depth value followed by a stabilization when the circularization is reached. However, with the current data, it is not possible to discard any of these configurations. { The second scenario would agree with the observed RV, but it would require a detailed stability analysis}. Other configurations could also be possible but more data are needed.

\subsection{Fast rotators: upper mass limits to {\it Kepler} planetary candidates (Group B)}
We have presented upper mass limits for six candidates with high rotational velocities. Because of this, our CAFE observations were unable to detect any RV variations below particular limits. These limits were used to set upper limits on the masses of the possible companions. { We found that the companions (if any) to four of these KOIs (namely KOI-0012, KOI-0131, KOI-0366, KOI-0625) have substellar masses  (i.e., $M< 0.08~M_{\odot}$), one of them (KOI-0366) having a maximum mass below the deuterium burning limit and thus being likely a true planet. On the contrary, we found that KOI-3728 is likely a false positive, since our analysis in Sect.~\S~\ref{sec:fastrotators} suggests the detection of a RV variation compatible with a stellar mass. However, the large rotational velocities prevents us from establishing a definite nature in both systems.}

\subsection{Unsolved cases (probable false positives,  Group C)}
We have analyzed two cases whose nature remains unsolved after our RV analysis. Although both of them show large RV variations { that strongly suggest that they are binaries}, we could not definitively unveil their nature because of the lack of enough phase coverage and the low number of data acquired. In both cases we have provided limits for the different parameters in the RV fit. 

{ In KOI-3890, we obtained a minimum mass for the transiting companion of $M_B> 0.097~M_{\odot}$, with the eccentricity of the orbit being $e>0.33$. In the case of KOI-1463, we obtain $M_B> 0.183~M_{\odot}$ and $e>0.18$. Thus, in both cases, the current data strongly suggests that the transiting companions are likely of stellar nature.}

%______________________________________________________________

\begin{acknowledgements}
      This research has been funded by Spanish grant AYA2012-38897-C02-01. J. Lillo-Box thanks the CSIC JAE-predoc program for the PhD fellowship. We also thank Calar Alto Observatory, both the open TAC and Spanish and MPIA GTO panel, for allocating our observing runs, and the CAHA staff for their effort and passion in their work, being very helpful during our visitor and service CAFE observations. PF and NCS acknowledge support by  Funda\c{c}\~ao para a Ci\^encia e a Tecnologia (FCT) through Investigador FCT contracts of reference IF/01037/2013 and IF/00169/2012, respectively, and POPH/FSE (EC) by FEDER funding through the program ``Programa Operacional de Factores de Competitividade - COMPETE''. We also acknowledge the support from the European Research Council/European Community under the FP7 through Starting Grant agreement number 239953. We acknowledge Gwena\"el Bou\'e for the useful discussion about the KOI-3853 system.
\end{acknowledgements}

%______________________________________________________________

\bibliographystyle{aa} % style aa.bst
\bibliography{biblio2} % your references Yourfile.bib

\clearpage

%______________________________________________________________

   % --- CAFE spectrum
   \begin{figure*}[HT]
   \centering
   \includegraphics[width=0.45\textwidth]{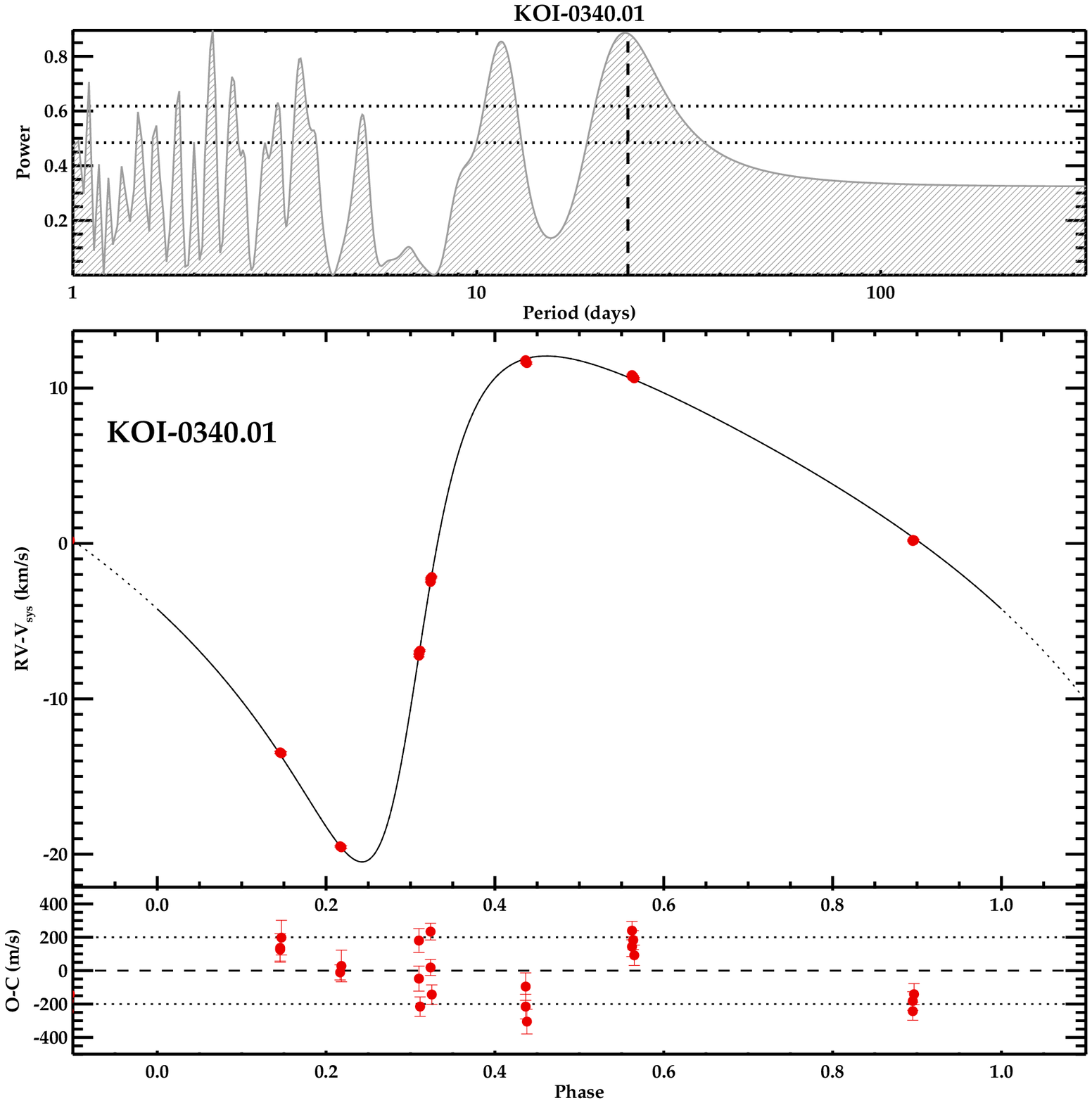}\includegraphics[width=0.45\textwidth]{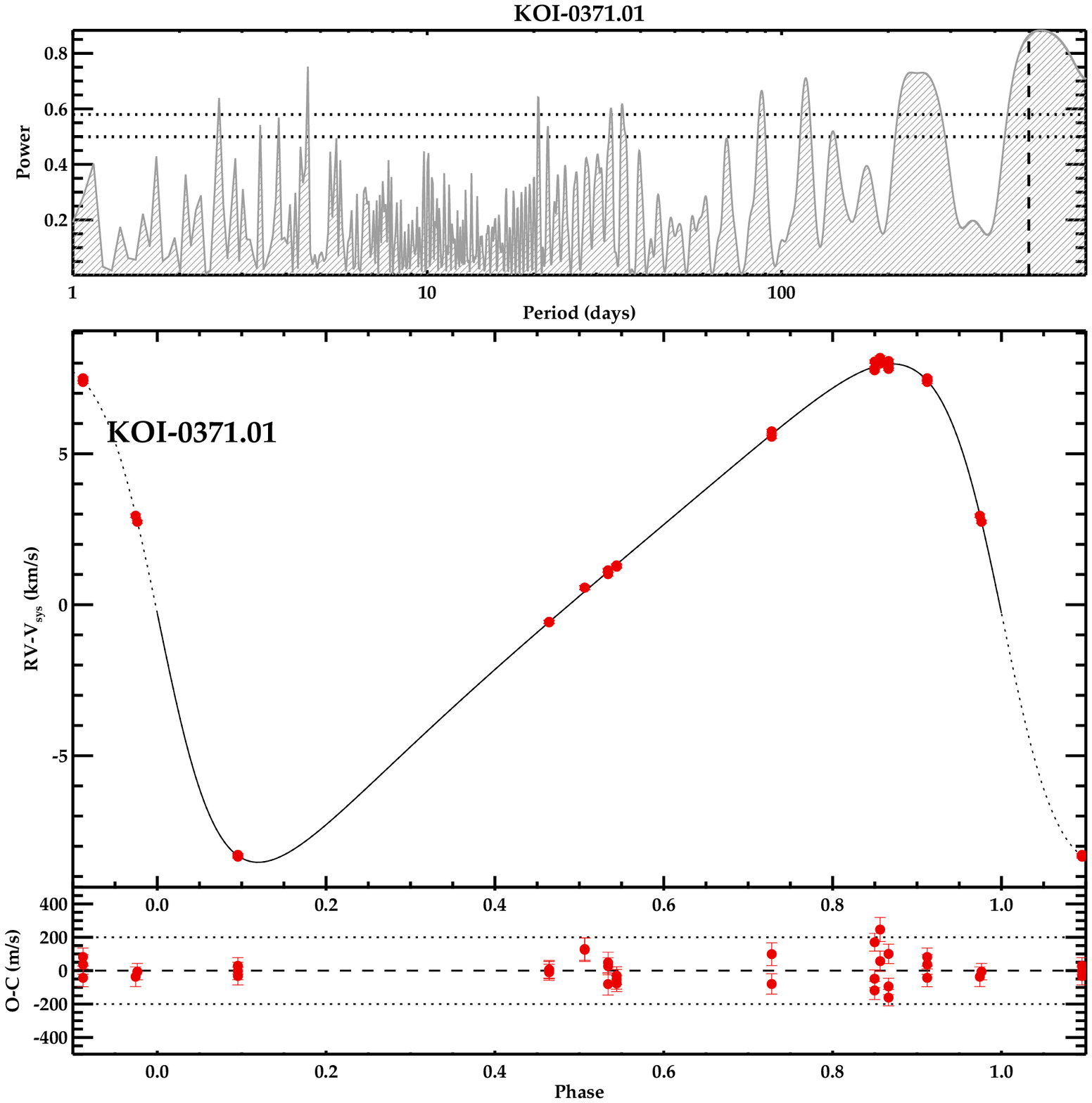} \includegraphics[width=0.45\textwidth]{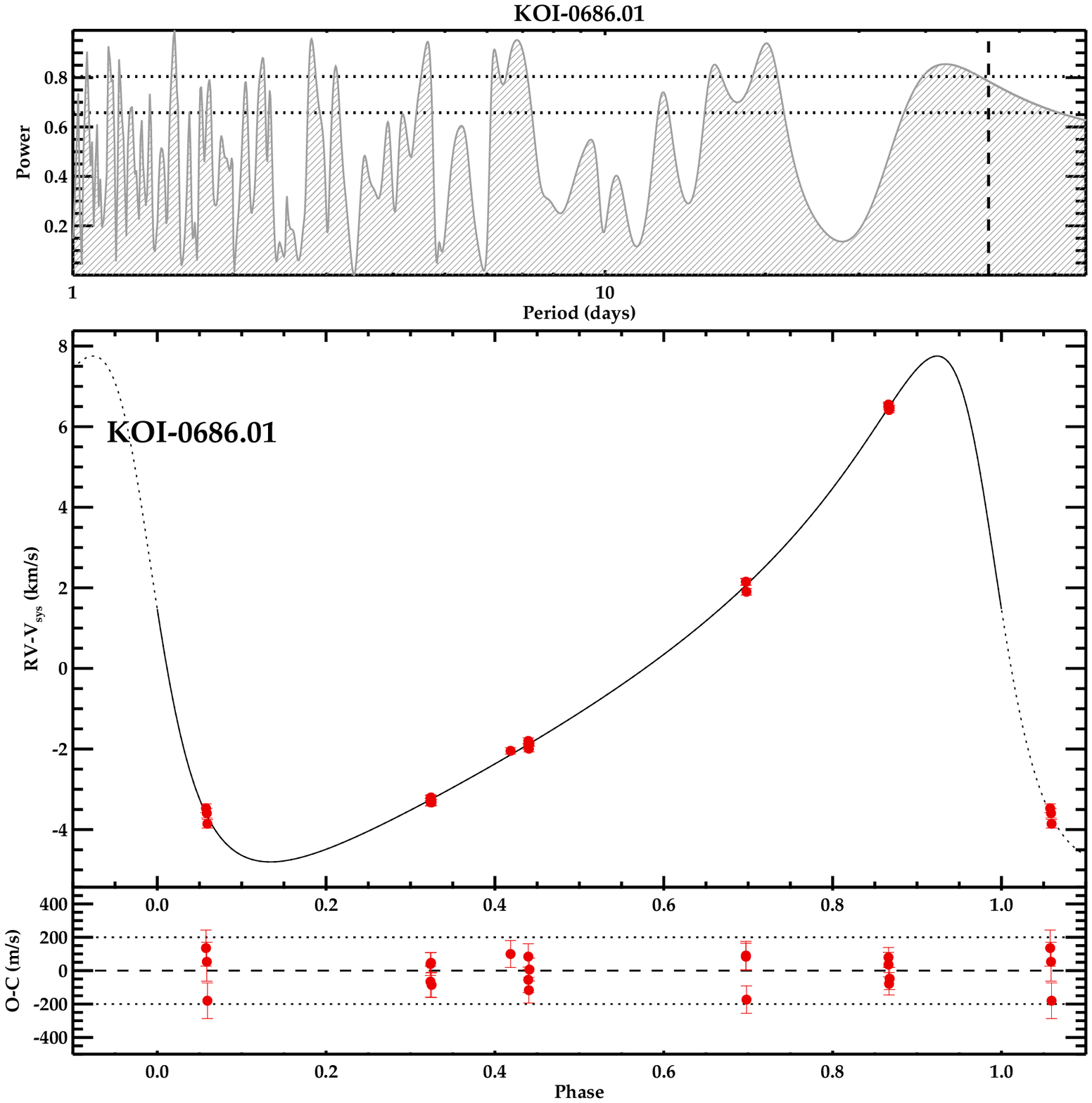}
  \includegraphics[width=0.45\textwidth]{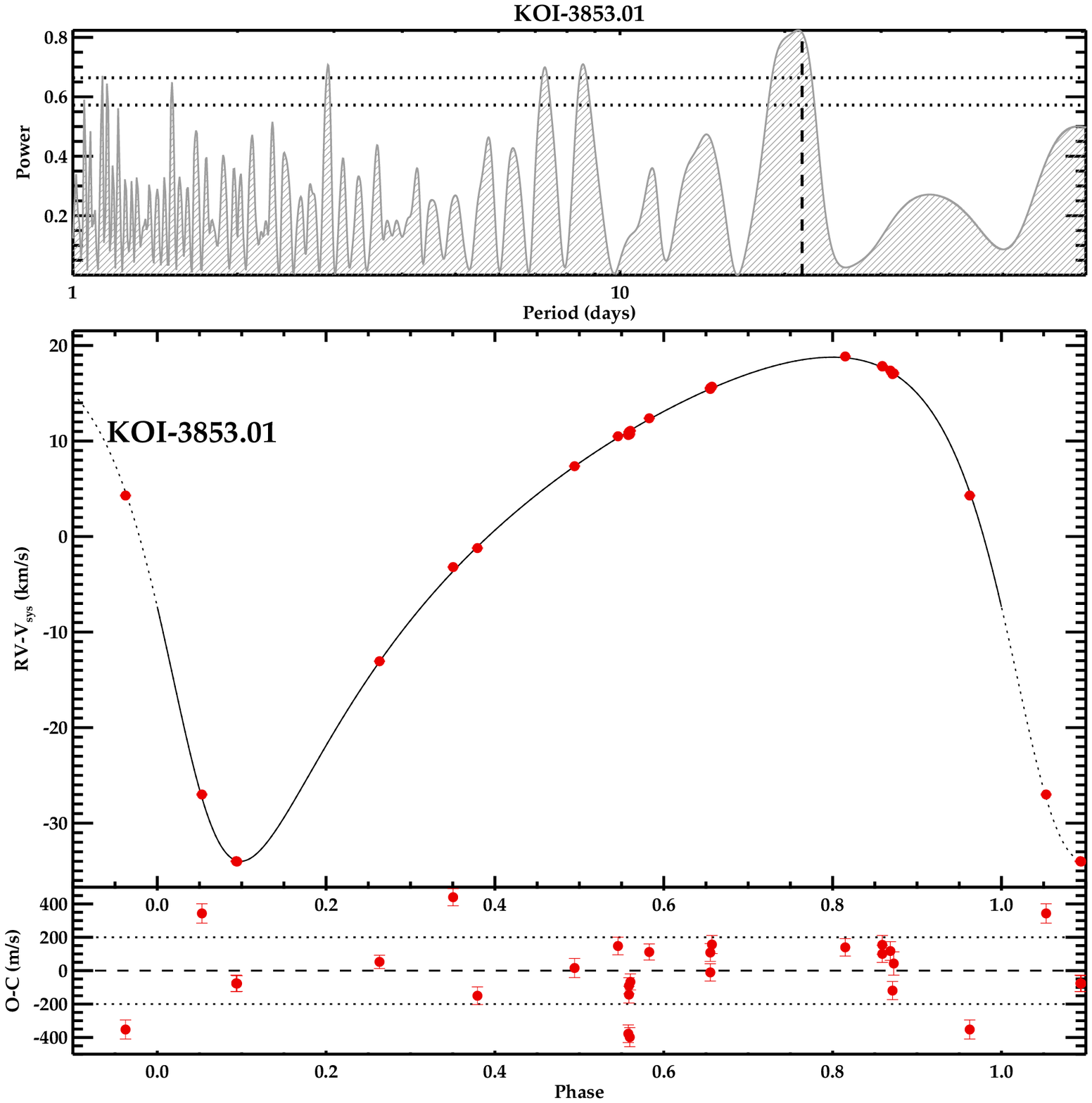}
         \caption{Radial velocity analysis of the detected binary systems. In each set of images for every KOI we show the periodogram of the radial velocity data (upper panel), the fit tot he data according to Sect.~\ref{sec:binaries} (middle panel), and the residuals of the fit (lower panel). }      
         \label{fig:rv}
   \end{figure*}

   % --- KOI-3853
   \begin{figure*}[HT]
   \centering
\includegraphics[width=0.9\textwidth]{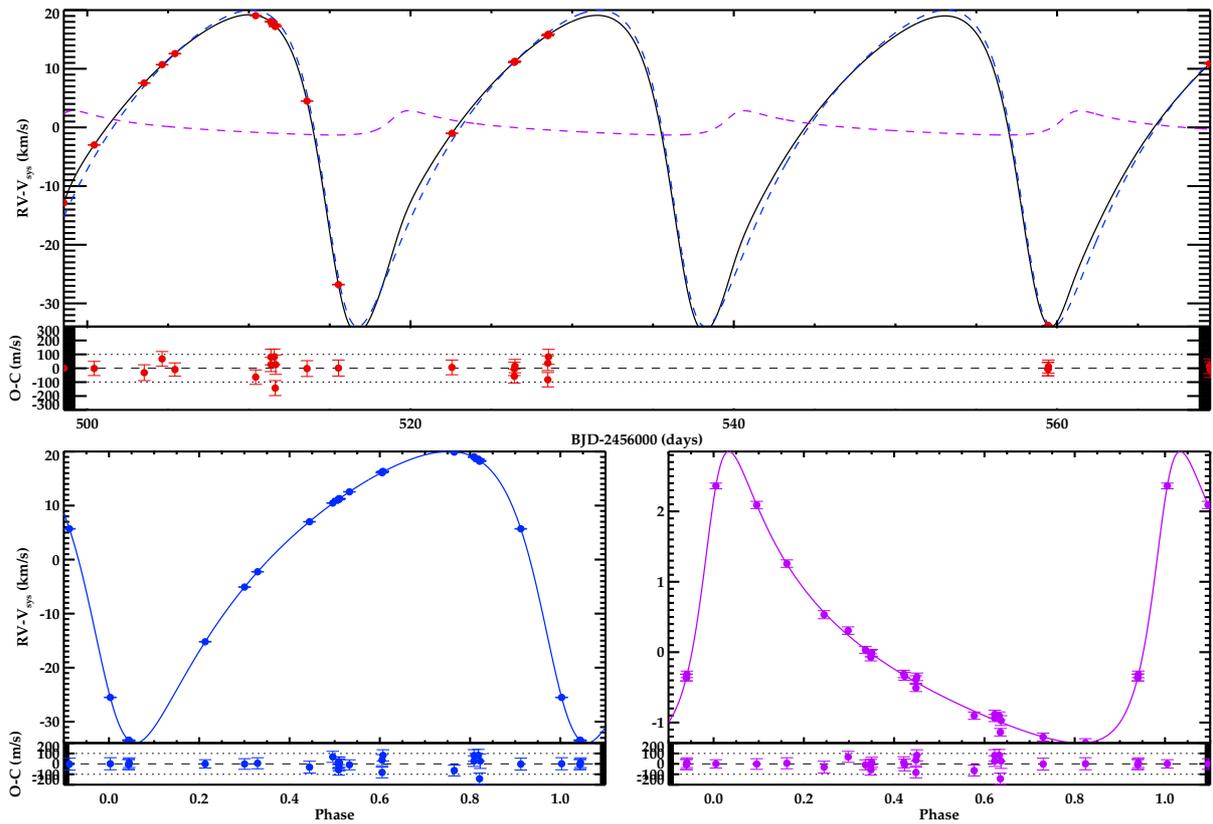}
         \caption{Radial velocity analysis of the detected components of KOI-3853.  Upper panel shows the RV solution for the whole system (black solid line) and the two contributions of the B component (dashed blue) and C component (dashed purple). Lower panels show the individual contributions of component B (left panel) and C (right panel). }
         \label{fig:koi3853}
   \end{figure*}

      % --- UNSOLVED
   \begin{figure*}[HT]
   \centering
\includegraphics[width=0.48\textwidth]{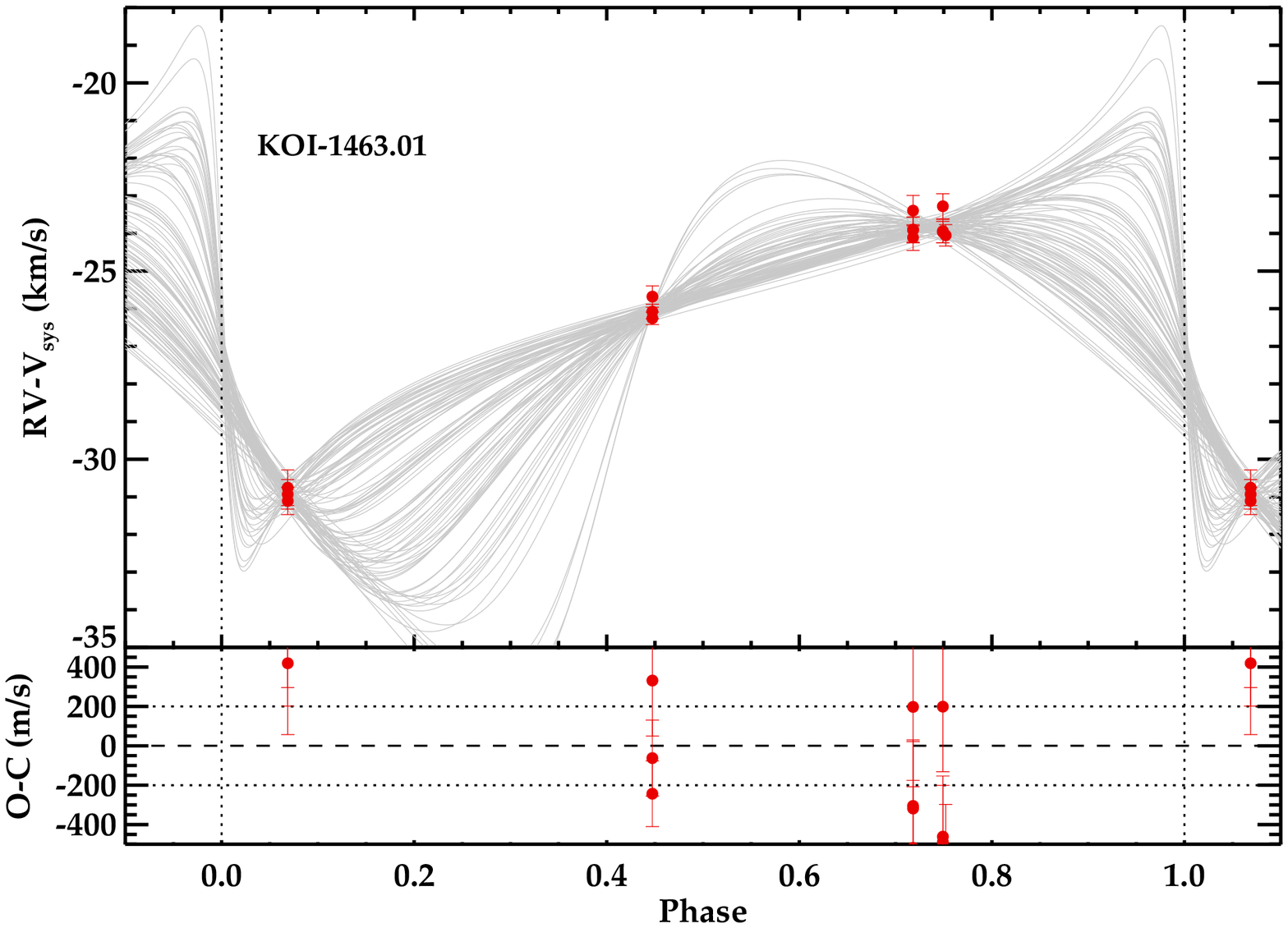}\hspace{0.1cm}\includegraphics[width=0.48\textwidth]{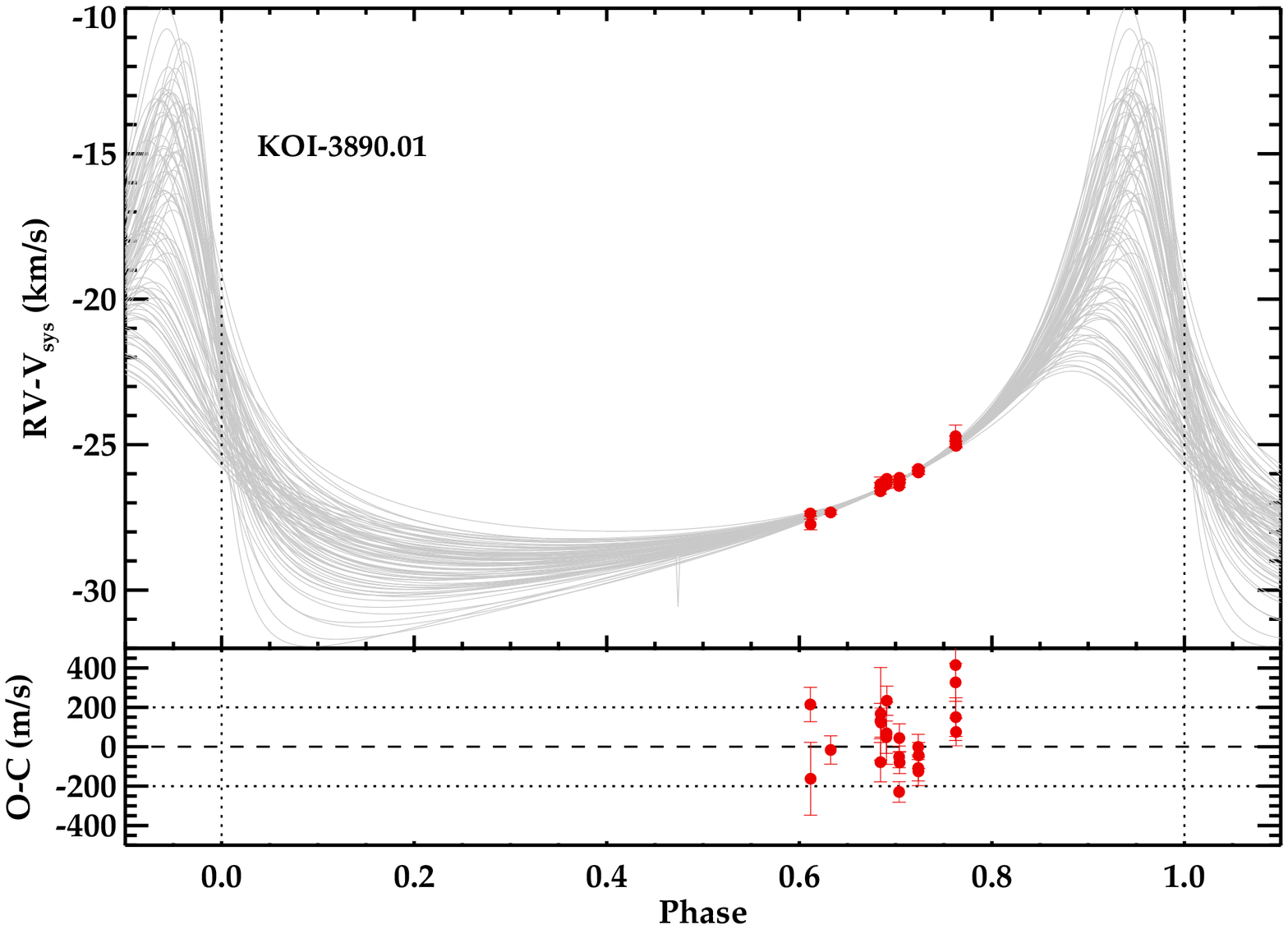}
         \caption{Radial velocity analysis of the unsolved systems (Group C, see Sect.~\ref{sec:unclear}). { We show one hundred random solutions accepted in our MCMC run (see Sect.~\ref{sec:unclear}). The residuals to the median solution are found in the lower panel}}
         \label{fig:rv2}
   \end{figure*}

\clearpage

   % --- LC KOI0340
   \begin{figure}[HT]
   \centering
   \includegraphics[width=0.5\textwidth]{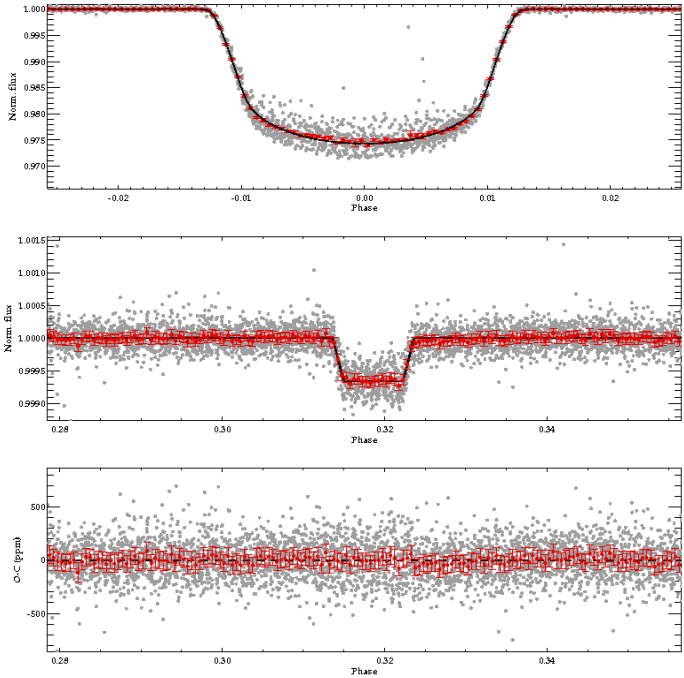}
   \caption{Dettrended {\it Kepler} light curve of KOI-0340. In all panels grey small circles represent the {\it Kepler} data and the red filled circles is a binning of the whole data with a bin size of 0.005 in phase. Upper panel shows a zoom to the primary transit and the fitted primary eclipse with a solid black line (see Sect.~\ref{sec:lc0340}). Middle panel shows the detected secondary eclipse and the fit obtained in Sect.~\ref{sec:lc0340} as a solid black line. The lower panel shows the residuals of the fit, having a standard deviation of 173 ppm in the original dataset and 36 ppm in the binned light-curve.}
         \label{fig:lc0340}
   \end{figure}

   % --- LC KOI3725
   \begin{figure}[HT]
   \centering
   \includegraphics[width=0.5\textwidth]{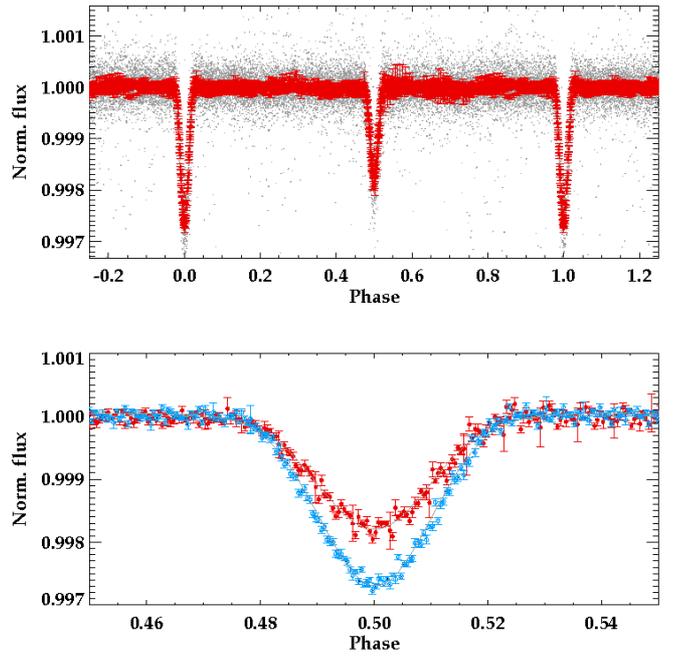}
   \caption{Dettrended {\it Kepler} light curve of KOI-3725 and phase-folded by using { the new calculated period (i.e., $P_{\rm orb}=3.1409940 \pm 0.0000008$ days),} twice that published by the {\it Kepler} team. The unequal depth of odd/even transits is a clear evidence of the binary nature of this object. Upper panel shows the  entire phase of the system. Lower panel shows a zoom to the odd and even transits (in different colors and symbols). The primary transit has been shifted to $\phi=0.5$ for comparison purposes. The transit model is shown as grey solid lines.}
         \label{fig:lc3725}
   \end{figure}

   % --- LC KOI3853
   \begin{figure}[HT]
   \centering
   \includegraphics[width=0.5\textwidth]{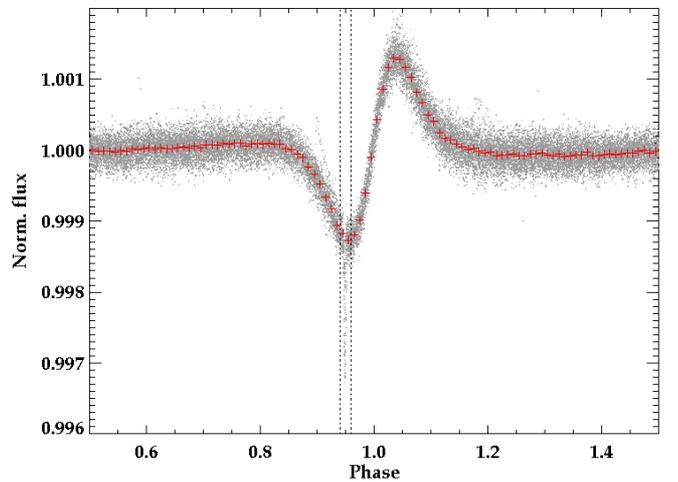}
   \caption{Dettrended {\it Kepler} light curve of KOI-3853 and phase-folded using the period published by the {\it Kepler} team. The upper panel shows the entire phase to show the heartbeat effect. The red line shows the spline function fitted to remove the effect. Vertical dotted lines show the region zoomed Fig.~\ref{fig:lc3853eclipses}. }
         \label{fig:lc3853}
   \end{figure}

      % --- LC KOI3853 eclipse depths
   \begin{figure}[h]
   \centering
   \includegraphics[width=0.5\textwidth]{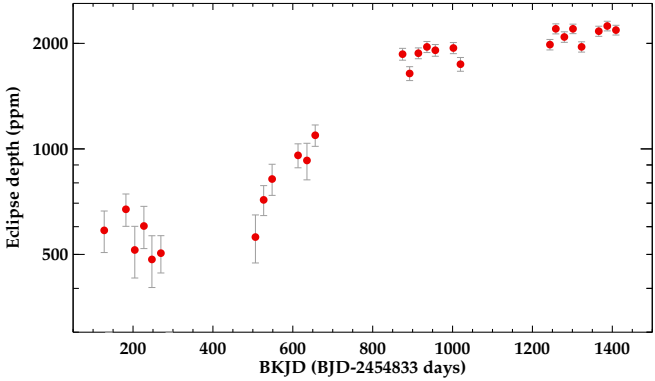}
   \caption{Variation of the eclipse depth of the KOI-3853 system in the time span of {\it Kepler} observations. }
         \label{fig:lc3853EclipseDepths}
   \end{figure}

   % --- LC KOI3853 eclipses
   \begin{figure*}[ht]
   \centering
   \includegraphics[width=0.75\textwidth]{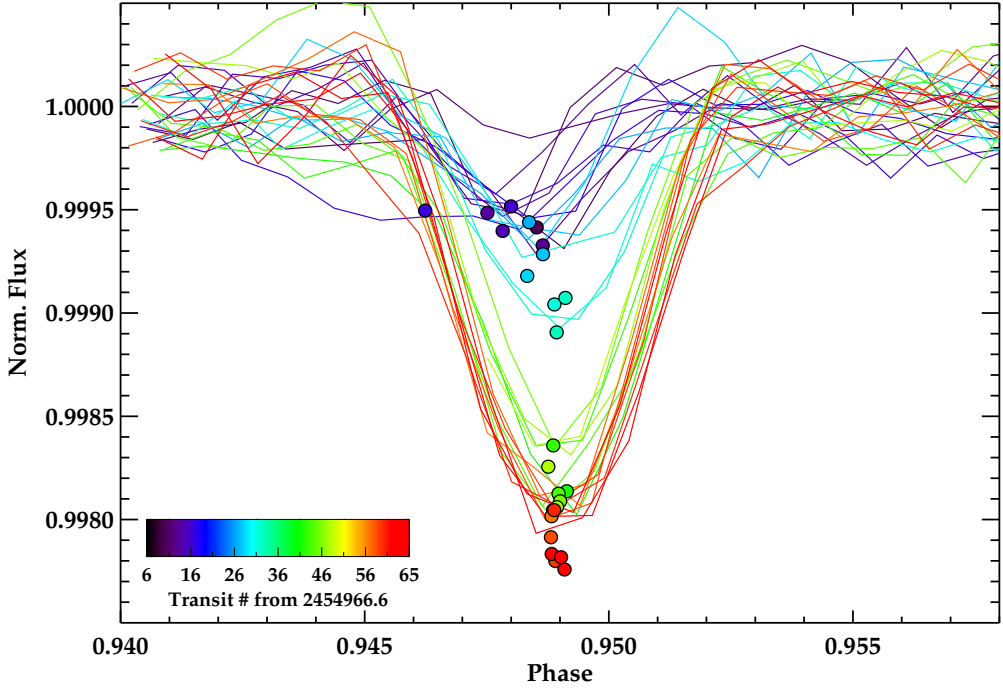}
   \caption{ Individual eclipses in the KOI-3853 system discussed in Sect. \S~\ref{sec:lc3853}. The color-code represents the transit number (equivalent to the julian date) since BJD $=2454966.6$ days. The filled circles show the location (timing and depth) of the peak of the eclipse for each individual transit. We have omitted time spans in which {\it Kepler} data is not available at the phases of interest. The light-curve has been firstly detrended  and then removed from the heartbeat effect as explained in Sect. \S~\ref{sec:lc3853}.  }
         \label{fig:lc3853eclipses}
   \end{figure*}
   
   % --- MASS-RADIUS
   \begin{figure}[ht]
   \centering
\includegraphics[width=0.5\textwidth]{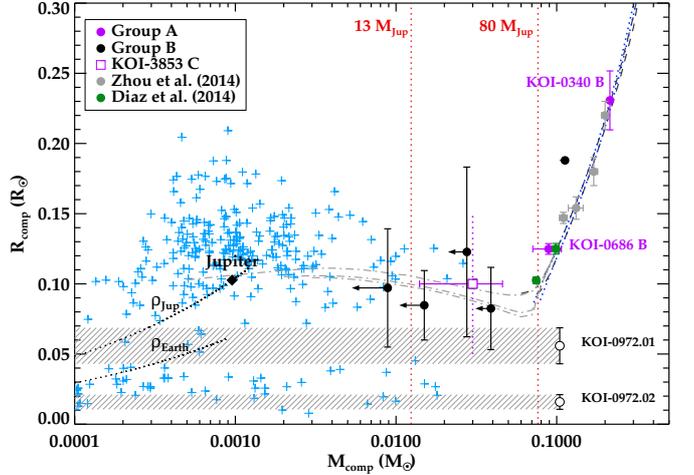}
         \caption{Mass-radius diagram of the companion objects with $M<0.3~M_{\odot}$ found to some of the analyzed KOIs in groups A and B. The open square shows the location of the possible third body in the KOI-3853 (i.e., KOI-3853 C), whith an {\it ad hoc} assumed radius of 0.1 $R_{\odot}$ (the dotted vertical error bar indicates our unknown about its actual size, { see Sect.~\S~\ref{sec:conclusions}}). We have used the isochrones of 1 Gyr and 5 Gyr (from top to bottom) from \cite{baraffe98} for solar metalicity (dotted dark blue line) and [Fe/H]=-0.5 (dashed light blue line). In the case of masses below 0.1 $M_{\odot}$, we have used the \cite{baraffe03} isochrones (dotted-dashed gray lines) of 1, 5, and 10 Gyr (from top to bottom). The arrows in the Group B KOIs indicate that their masses are just upper limits. The dashed regions illustrates the range of masses allowed for the KOI-0972 planet candidates (see Sect.~\S~\ref{sec:conclusions}). { For reference, we have plotted as light blue plus signs the known planets with provided masses and radius published in the exoplanet encyclopedia (www.exoplanet.eu).} }
         \label{fig:MR}
   \end{figure}

\pagebreak
\clearpage
\newpage

\appendix

\section{Radial velocity tables \label{app:A}}

\begin{table*}[htbf]
\setlength{\extrarowheight}{2pt}
\scriptsize
\caption{KOI-0012: measured radial velocities. The S/N in Tables A.1 to A.13 is calculated as the median signal over the measured scatter of a continuum region close to 5500 \AA.}
\label{tab:KOI-0012}
\centering 
\begin{tabular}{ccc|ccc|ccc}     % 7 columns 
\hline\hline
 Julian Date & $\overline{\rm S/N}$ &  RV & Julian Date & $\overline{\rm S/N}$ &  RV & Julian Date & $\overline{\rm S/N}$ &  RV       \\ 
(days)-2456000 & & (km/s) & (days)-2456000 & & (km/s) & (days)-2456000 & & (km/s) \\ \hline
           66.442058  &       14.4  &                  $-18.3^{+6.5}_{-4.7}$ &           72.463525  &       15.5  &                  $-18.1^{+3.1}_{-2.7}$ &           87.634276  &       15.5  &                    $-18.5^{+4.6}_{-1.2}$ \\
           66.454881  &       24.0  &                   $-9.4^{+5.1}_{-4.7}$ &           76.518732  &       18.1  &                  $-19.2^{+2.4}_{-3.4}$ &           90.630156  &       21.5  &                    $-18.9^{+3.3}_{-4.4}$ \\
           66.467705  &       19.0  &                   $-9.6^{+5.6}_{-5.7}$ &           76.533549  &       34.2  &                  $-17.5^{+1.6}_{-3.4}$ &          115.379299  &       21.0  &                    $-16.5^{+2.8}_{-4.5}$ \\
           69.439270  &       15.0  &                  $-18.4^{+7.1}_{-6.2}$ &           76.548365  &       28.8  &                  $-17.6^{+1.6}_{-1.8}$ &          115.391505  &       29.9  &                    $-18.0^{+2.0}_{-1.5}$ \\
           69.451609  &       21.1  &                  $-23.0^{+7.2}_{-1.4}$ &           77.602751  &       22.5  &                  $-21.4^{+3.6}_{-1.4}$ &          115.403712  &       21.1  &                    $-20.7^{+3.3}_{-1.2}$ \\
           69.463949  &       14.7  &                  $-19.5^{+6.7}_{-2.5}$ &           77.615244  &       30.5  &                  $-19.0^{+2.1}_{-2.8}$ &          117.381785  &       22.3  &                    $-18.7^{+2.3}_{-1.5}$ \\
           70.434806  &       24.8  &                  $-18.1^{+1.7}_{-2.2}$ &           77.627738  &       20.5  &                  $-13.5^{+3.8}_{-3.5}$ &          117.393940  &       27.0  &                    $-14.3^{+3.6}_{-1.3}$ \\
           70.447044  &       31.4  &                  $-20.6^{+1.8}_{-1.3}$ &           78.457637  &       23.3  &                  $-19.3^{+3.5}_{-2.5}$ &          117.406095  &       15.0  &                    $-20.9^{+8.6}_{-1.7}$ \\
           70.459281  &       19.1  &                  $-21.8^{+2.8}_{-1.3}$ &           78.472721  &       29.5  &                  $-19.0^{+1.9}_{-2.3}$ &          129.381270  &       17.0  &                    $-21.9^{+4.8}_{-1.7}$ \\
           71.431319  &       23.0  &                  $-18.9^{+2.7}_{-1.3}$ &           78.487805  &       17.8  &                  $-14.8^{+3.9}_{-2.7}$ &          129.393473  &       27.8  &                    $-20.0^{+2.8}_{-2.5}$ \\
           71.456236  &       25.8  &                  $-19.2^{+1.5}_{-1.6}$ &           80.406010  &       19.8  &                  $-21.3^{+4.0}_{-2.0}$ &          129.405676  &       21.9  &                    $-18.7^{+2.0}_{-1.5}$ \\
           71.532211  &       53.8  &                  $-16.6^{+1.4}_{-2.8}$ &           80.418276  &       31.2  &                  $-19.8^{+3.0}_{-7.5}$ &          180.442531  &       21.3  &                    $-21.2^{+3.9}_{-1.7}$ \\
           71.608528  &       30.3  &                  $-16.6^{+1.6}_{-2.1}$ &           80.430543  &       24.1  &                  $-17.7^{+2.5}_{-4.6}$ &          180.464569  &       23.0  &                    $-19.5^{+1.9}_{-2.7}$ \\
           71.632761  &       29.8  &                  $-19.4^{+1.7}_{-1.4}$ &           82.602553  &       20.3  &                  $-19.6^{+1.7}_{-1.7}$ &          180.464653  &       41.1  &                    $-19.6^{+1.8}_{-2.1}$ \\
           72.439229  &       23.2  &                  $-17.9^{+1.9}_{-3.1}$ &           82.614673  &       26.2  &                  $-20.2^{+1.8}_{-2.1}$ &          180.486860  &       25.9  &                    $-20.2^{+1.4}_{-1.7}$ \\
           72.451377  &       27.8  &                  $-17.7^{+1.8}_{-3.2}$ &           82.626794  &       16.6  &                  $-16.7^{+3.3}_{-1.7}$ &                      &             &                                          \\
\hline 
\end{tabular}
%\tablefoot{
%\tablefoottext{a}{}
%}
\end{table*}

\begin{table*}[htbf]
\setlength{\extrarowheight}{2pt}
\scriptsize
\caption{KOI-0131: measured radial velocities}
\label{tab:KOI-0131}
\centering 
\begin{tabular}{ccc|ccc|ccc}     % 7 columns 
\hline\hline
 Julian Date & $\overline{\rm S/N}$ &  RV & Julian Date & $\overline{\rm S/N}$ &  RV & Julian Date & $\overline{\rm S/N}$ &  RV       \\ 
(days)-2456000 & & (km/s) & (days)-2456000 & & (km/s) & (days)-2456000 & & (km/s) \\ \hline
          111.576836  &        7.0  &                   $-6.2^{+1.7}_{-1.4}$ &          120.636824  &        6.5  &                   $-5.3^{+3.1}_{-3.4}$ &          497.517894  &        4.2  &                     $-8.7^{+4.4}_{-8.8}$ \\
          115.458790  &        9.1  &                   $-7.8^{+1.6}_{-1.1}$ &          418.567422  &        6.5  &                  $-12.1^{+9.3}_{-2.2}$ &          514.364588  &        4.6  &                     $-8.1^{+8.9}_{-2.2}$ \\
          115.493531  &        7.4  &                  $-10.7^{+2.8}_{-2.3}$ &          418.599802  &        7.6  &                 $-11.8^{+2.8}_{-0.11}$ &          514.401807  &       11.2  &                    $-21.5^{+5.3}_{-2.9}$ \\
          115.493595  &       14.0  &                  $-11.6^{+4.4}_{-1.8}$ &          418.632175  &       12.2  &                  $-10.5^{+2.9}_{-4.8}$ &          514.402164  &        7.0  &                     $-8.7^{+2.1}_{-2.8}$ \\
          115.528462  &        8.0  &                   $-8.1^{+2.4}_{-2.0}$ &          436.501911  &        7.1  &                $159.32^{+0.12}_{-4.4}$ &          514.438668  &        5.2  &                      $3.3^{+6.2}_{-3.6}$ \\
          119.544606  &        4.5  &                   $-8.4^{+2.6}_{-1.3}$ &          436.538388  &        6.5  &                    $0.0^{+4.8}_{-5.1}$ &          529.484818  &        8.7  &                    $-12.1^{+4.0}_{-2.3}$ \\
          120.530916  &        7.0  &                   $-8.7^{+1.4}_{-3.3}$ &          436.573185  &        5.2  &                  $-10.0^{+4.7}_{-2.6}$ &          529.521132  &       12.7  &                    $-13.4^{+2.6}_{-4.0}$ \\
          120.566218  &        6.2  &                   $-5.0^{+4.1}_{-3.0}$ &          497.439970  &        5.2  &                   $-3.5^{+6.5}_{-4.2}$ &          529.521166  &        7.2  &                      $-6.1^{+3.4}_{-11}$ \\
          120.583937  &       13.4  &                   $-7.9^{+2.9}_{-2.9}$ &          497.475725  &        6.2  &                    $-5.3^{+3.1}_{-11}$ &          529.557411  &        4.5  &                    $-12.9^{+3.4}_{-1.8}$ \\
          120.601789  &        6.2  &                   $-8.5^{+3.2}_{-4.6}$ &          497.477863  &        9.5  &                  $-11.4^{+5.0}_{-2.5}$ &                      &             &                                          \\
\hline 
\end{tabular}
%\tablefoot{
%\tablefoottext{a}{}
%}
\end{table*}

\begin{table*}[htbf]
\setlength{\extrarowheight}{2pt}
\scriptsize
\caption{KOI-0340: measured radial velocities}
\label{tab:KOI-0340}
\centering 
\begin{tabular}{ccc|ccc|ccc}     % 7 columns 
\hline\hline
 Julian Date & $\overline{\rm S/N}$ &  RV & Julian Date & $\overline{\rm S/N}$ &  RV & Julian Date & $\overline{\rm S/N}$ &  RV       \\ 
(days)-2456000 & & (km/s) & (days)-2456000 & & (km/s) & (days)-2456000 & & (km/s) \\ \hline
          111.395672  &       13.8  &           $-103.499^{+0.045}_{-0.045}$ &          116.633232  &        9.5  &            $-72.401^{+0.074}_{-0.057}$ &          127.489427  &       15.8  &              $-83.806^{+0.063}_{-0.058}$ \\
          111.431211  &       12.8  &           $-103.559^{+0.095}_{-0.095}$ &          119.579846  &       11.6  &            $-73.268^{+0.059}_{-0.061}$ &          133.386589  &       12.5  &              $-97.467^{+0.069}_{-0.071}$ \\
          113.605011  &       12.5  &            $-91.229^{+0.075}_{-0.075}$ &          119.612757  &       19.6  &            $-73.172^{+0.055}_{-0.054}$ &          133.404004  &       15.2  &              $-97.457^{+0.085}_{-0.074}$ \\
          113.621494  &       17.8  &            $-91.001^{+0.071}_{-0.058}$ &          119.614455  &       12.4  &            $-73.262^{+0.057}_{-0.063}$ &          133.421419  &        8.1  &                $-97.52^{+0.10}_{-0.055}$ \\
          113.637977  &       12.5  &            $-90.909^{+0.058}_{-0.082}$ &          119.643968  &        9.2  &            $-73.381^{+0.061}_{-0.069}$ &          137.604302  &        9.5  &              $-86.483^{+0.048}_{-0.056}$ \\
          116.598403  &        7.8  &            $-72.213^{+0.082}_{-0.074}$ &          127.454588  &       15.5  &             $-83.850^{+0.054}_{-0.10}$ &          137.621876  &       14.1  &              $-86.268^{+0.050}_{-0.085}$ \\
          116.615818  &       12.6  &            $-72.333^{+0.074}_{-0.059}$ &          127.472007  &       21.9  &            $-83.789^{+0.056}_{-0.048}$ &          137.639450  &        9.9  &              $-86.171^{+0.058}_{-0.050}$ \\
\hline 
\end{tabular}
%\tablefoot{
%\tablefoottext{a}{}
%}
\end{table*}

\begin{table*}[htbf]
\setlength{\extrarowheight}{2pt}
\scriptsize
\caption{KOI-0366: measured radial velocities}
\label{tab:KOI-0366}
\centering 
\begin{tabular}{ccc|ccc|ccc}     % 7 columns 
\hline\hline
 Julian Date & $\overline{\rm S/N}$ &  RV & Julian Date & $\overline{\rm S/N}$ &  RV & Julian Date & $\overline{\rm S/N}$ &  RV       \\ 
(days)-2456000 & & (km/s) & (days)-2456000 & & (km/s) & (days)-2456000 & & (km/s) \\ \hline
           66.494606  &       15.4  &                    $6.8^{+1.2}_{-1.0}$ &           82.469641  &       15.4  &                   $8.3^{+1.3}_{-0.41}$ &          112.644079  &       15.4  &                   $7.45^{+0.52}_{-0.46}$ \\
           69.490469  &       15.4  &                    $7.7^{+1.8}_{-1.6}$ &           82.487003  &       15.4  &                 $7.94^{+0.78}_{-0.39}$ &          118.374379  &       15.4  &                   $8.19^{+0.47}_{-0.31}$ \\
           71.558576  &       15.4  &                 $7.85^{+0.37}_{-0.28}$ &           82.504365  &       15.4  &                 $7.77^{+0.84}_{-0.25}$ &          118.398555  &       15.4  &                   $8.79^{+0.40}_{-0.24}$ \\
           71.570735  &       15.4  &                 $7.15^{+0.38}_{-0.31}$ &           92.591566  &       15.4  &                 $9.07^{+0.40}_{-0.53}$ &          118.398689  &       15.4  &                   $9.00^{+0.29}_{-0.53}$ \\
           71.582894  &       15.4  &                 $7.50^{+0.40}_{-0.27}$ &           92.615777  &       15.4  &                 $9.11^{+0.38}_{-0.71}$ &          118.423132  &       15.4  &                   $8.06^{+0.38}_{-0.33}$ \\
           72.599880  &       15.4  &                 $9.06^{+0.32}_{-0.30}$ &           92.615814  &       15.4  &                 $9.06^{+0.33}_{-0.30}$ &          126.369758  &       15.4  &                   $7.87^{+0.51}_{-0.61}$ \\
           72.612031  &       15.4  &                 $8.80^{+0.31}_{-0.77}$ &           92.640099  &       15.4  &                 $8.93^{+0.44}_{-0.45}$ &          126.382150  &       15.4  &                   $7.93^{+0.55}_{-0.28}$ \\
           72.624182  &       15.4  &                 $8.37^{+0.36}_{-0.72}$ &          101.543893  &       15.4  &                 $7.17^{+0.44}_{-0.41}$ &          126.394542  &       15.4  &                   $9.36^{+0.64}_{-0.26}$ \\
           76.440887  &       15.4  &                 $9.00^{+0.93}_{-0.42}$ &          101.561229  &       15.4  &                 $8.64^{+0.34}_{-0.34}$ &          130.380236  &       15.4  &                   $7.95^{+0.36}_{-0.51}$ \\
           76.466954  &       15.4  &                 $9.54^{+0.65}_{-0.46}$ &          101.578566  &       15.4  &                 $8.69^{+0.30}_{-0.30}$ &          130.392427  &       15.4  &                   $8.29^{+0.31}_{-0.42}$ \\
           76.467735  &       15.4  &                 $9.95^{+0.88}_{-0.29}$ &          110.385278  &       15.4  &                 $7.93^{+0.62}_{-0.39}$ &          130.404619  &       15.4  &                   $9.46^{+0.32}_{-0.24}$ \\
           76.492240  &       15.4  &                  $8.27^{+0.50}_{-1.1}$ &          110.397526  &       15.4  &                 $7.46^{+0.65}_{-0.56}$ &          139.381895  &       15.4  &                   $7.75^{+0.57}_{-0.51}$ \\
           79.543631  &       15.4  &                 $9.61^{+0.55}_{-0.68}$ &          110.409774  &       15.4  &                 $7.57^{+0.88}_{-0.32}$ &          139.394228  &       15.4  &                   $8.35^{+0.51}_{-0.25}$ \\
           79.555913  &       15.4  &                 $8.67^{+0.39}_{-0.32}$ &          112.619832  &       15.4  &                 $8.67^{+0.34}_{-0.27}$ &          139.406562  &       15.4  &                   $8.41^{+0.53}_{-0.44}$ \\
           79.568194  &       15.4  &                 $8.59^{+0.34}_{-0.33}$ &          112.631955  &       15.4  &                 $7.36^{+0.49}_{-0.47}$ &                      &             &                                          \\
\hline 
\end{tabular}
%\tablefoot{
%\tablefoottext{a}{}
%}
\end{table*} 

\begin{table*}[htbf]
\setlength{\extrarowheight}{2pt}
\scriptsize
\caption{KOI-0371: measured radial velocities}
\label{tab:KOI-0371}
\centering 
\begin{tabular}{ccc|ccc|ccc}     % 7 columns 
\hline\hline
 Julian Date & $\overline{\rm S/N}$ &  RV & Julian Date & $\overline{\rm S/N}$ &  RV & Julian Date & $\overline{\rm S/N}$ &  RV       \\ 
(days)-2456000 & & (km/s) & (days)-2456000 & & (km/s) & (days)-2456000 & & (km/s) \\ \hline
          121.454974  &       11.8  &            $-65.169^{+0.052}_{-0.052}$ &          505.601217  &        6.5  &            $-53.219^{+0.049}_{-0.053}$ &          803.627383  &        9.6  &              $-65.601^{+0.055}_{-0.059}$ \\
          121.467317  &       17.7  &            $-65.138^{+0.049}_{-0.053}$ &          505.612199  &        9.8  &            $-53.152^{+0.049}_{-0.059}$ &          824.620384  &        5.5  &              $-64.962^{+0.067}_{-0.048}$ \\
          121.479659  &       13.2  &            $-65.200^{+0.053}_{-0.061}$ &          505.623181  &        7.5  &            $-52.957^{+0.058}_{-0.059}$ &          824.642337  &        5.2  &              $-64.965^{+0.070}_{-0.049}$ \\
          436.604913  &       13.7  &            $-58.579^{+0.061}_{-0.068}$ &          528.419975  &       10.2  &            $-53.651^{+0.053}_{-0.049}$ &          838.397849  &        9.2  &              $-64.515^{+0.066}_{-0.053}$ \\
          436.629188  &       13.7  &            $-58.399^{+0.068}_{-0.054}$ &          528.438254  &       18.5  &            $-53.526^{+0.054}_{-0.050}$ &          838.416113  &       14.1  &              $-64.386^{+0.062}_{-0.049}$ \\
          497.381256  &        9.6  &            $-53.264^{+0.054}_{-0.053}$ &          528.456533  &       15.7  &            $-53.573^{+0.059}_{-0.055}$ &          838.434376  &       10.2  &              $-64.406^{+0.051}_{-0.054}$ \\
          497.394856  &       15.2  &            $-53.194^{+0.053}_{-0.072}$ &          559.495749  &       11.7  &            $-58.079^{+0.059}_{-0.067}$ &          843.450061  &       11.3  &              $-64.246^{+0.059}_{-0.048}$ \\
          497.408455  &       10.8  &            $-52.975^{+0.053}_{-0.060}$ &          560.496212  &       14.5  &            $-58.283^{+0.049}_{-0.070}$ &          843.466246  &       16.1  &              $-64.224^{+0.053}_{-0.062}$ \\
          500.645274  &        5.6  &            $-52.852^{+0.072}_{-0.049}$ &          803.605301  &        9.3  &            $-65.603^{+0.050}_{-0.066}$ &          843.482431  &       11.2  &              $-64.270^{+0.048}_{-0.053}$ \\
          500.669675  &        9.0  &            $-53.041^{+0.060}_{-0.058}$ &          803.616342  &       13.5  &            $-65.618^{+0.048}_{-0.051}$ &                      &             &                                          \\
\hline 
\end{tabular}
%\tablefoot{
%\tablefoottext{a}{}
%}
\end{table*} 

\begin{table*}[htbf]
\setlength{\extrarowheight}{2pt}
\scriptsize
\caption{KOI-0625: measured radial velocities}
\label{tab:KOI-0625}
\centering 
\begin{tabular}{ccc|ccc|ccc}     % 7 columns 
\hline\hline
 Julian Date & $\overline{\rm S/N}$ &  RV & Julian Date & $\overline{\rm S/N}$ &  RV & Julian Date & $\overline{\rm S/N}$ &  RV       \\ 
(days)-2456000 & & (km/s) & (days)-2456000 & & (km/s) & (days)-2456000 & & (km/s) \\ \hline
           71.489198  &       11.6  &               $-25.97^{+0.55}_{-0.45}$ &           88.573506  &        1.7  &                 $-28.8^{+4.2}_{-0.89}$ &           94.589328  &        9.8  &                  $-25.77^{+0.98}_{-1.6}$ \\
           71.506591  &       16.5  &               $-25.80^{+0.62}_{-0.59}$ &           88.592027  &        9.8  &               $-26.27^{+0.74}_{-0.67}$ &          102.432371  &        9.3  &                   $-24.9^{+1.1}_{-0.56}$ \\
           71.523984  &       11.6  &               $-26.13^{+0.65}_{-0.64}$ &           88.597524  &       13.0  &                 $-23.8^{+1.9}_{-0.99}$ &          102.469438  &        9.1  &                 $-26.92^{+0.99}_{-0.54}$ \\
           72.496141  &       10.7  &                $-26.39^{+0.79}_{-1.1}$ &           88.627038  &        7.8  &                  $-24.6^{+1.7}_{-1.8}$ &          102.469462  &       15.6  &                   $-31.2^{+2.5}_{-0.72}$ \\
           72.513475  &       15.2  &               $-25.29^{+0.84}_{-0.55}$ &           91.459181  &        8.3  &                 $-25.9^{+1.4}_{-0.85}$ &          102.506577  &        7.8  &                   $-24.3^{+1.8}_{-0.45}$ \\
           72.530808  &       10.7  &                 $-25.5^{+1.3}_{-0.49}$ &           91.493991  &       15.7  &                 $-25.1^{+1.3}_{-0.75}$ &          121.559629  &        8.1  &                   $-25.2^{+1.1}_{-0.73}$ \\
           77.443828  &       11.6  &               $-25.57^{+0.69}_{-0.89}$ &           91.494063  &        9.0  &                 $-26.1^{+1.2}_{-0.84}$ &          121.594577  &       13.7  &                    $-24.1^{+1.2}_{-3.8}$ \\
           77.462844  &       16.5  &               $-26.31^{+0.59}_{-0.62}$ &           91.528730  &        9.5  &               $-26.64^{+0.95}_{-0.90}$ &          121.594628  &        7.6  &                    $-26.5^{+1.6}_{-2.3}$ \\
           77.481860  &       11.8  &                $-26.66^{+0.65}_{-5.3}$ &           92.525138  &       11.2  &               $-25.37^{+0.91}_{-0.95}$ &          121.629474  &        7.8  &                    $-24.7^{+1.3}_{-1.9}$ \\
           79.476176  &        9.0  &                 $-25.5^{+1.1}_{-0.92}$ &           92.542705  &       16.1  &                $-26.08^{+0.97}_{-1.9}$ &          127.526771  &        9.7  &                  $-25.03^{+0.99}_{-1.1}$ \\
           79.493735  &       14.0  &                $-27.39^{+0.96}_{-3.5}$ &           92.560272  &       10.9  &                $-25.77^{+0.86}_{-1.1}$ &          127.561701  &        9.0  &                    $-26.1^{+1.0}_{-1.1}$ \\
           79.511293  &       10.4  &                $-25.08^{+0.63}_{-2.4}$ &           93.453444  &       10.2  &                  $-25.0^{+1.2}_{-1.4}$ &          127.561732  &       16.5  &                  $-25.92^{+0.89}_{-1.9}$ \\
           80.462852  &        8.2  &                  $-29.7^{+5.4}_{-3.5}$ &           93.471006  &       14.6  &                  $-24.2^{+1.7}_{-1.2}$ &          127.596724  &        8.9  &                 $-26.22^{+0.61}_{-0.92}$ \\
           80.487557  &       13.2  &                 $-23.6^{+4.4}_{-0.62}$ &           93.488568  &       10.3  &                  $-27.6^{+1.7}_{-1.0}$ &          137.461469  &        8.5  &                    $-25.1^{+1.1}_{-2.4}$ \\
           80.512261  &       10.1  &                  $-25.4^{+1.0}_{-1.4}$ &           94.519592  &       10.9  &               $-26.06^{+0.94}_{-0.85}$ &          137.497308  &        7.9  &                    $-23.3^{+1.9}_{-1.3}$ \\
           87.567872  &        7.5  &                  $-25.5^{+4.1}_{-1.2}$ &           94.554529  &       18.6  &                 $-25.0^{+1.1}_{-0.41}$ &          137.498740  &       15.2  &                   $-27.1^{+1.9}_{-0.72}$ \\
           87.585564  &       10.0  &                 $-25.7^{+2.8}_{-0.95}$ &           94.554666  &       10.3  &                $-25.29^{+0.89}_{-1.1}$ &          137.537443  &        9.5  &                 $-26.52^{+0.64}_{-1.00}$ \\
           87.603257  &        6.2  &                 $-24.7^{+2.7}_{-0.71}$ &                      &             &                                        &                      &             &                                          \\
\hline 
\end{tabular}
%\tablefoot{
%\tablefoottext{a}{}
%}
\end{table*} 

\begin{table*}[htbf]
\setlength{\extrarowheight}{2pt}
\scriptsize
\caption{KOI-0686: measured radial velocities}
\label{tab:KOI-0686}
\centering 
\begin{tabular}{ccc|ccc|ccc}     % 7 columns 
\hline\hline
 Julian Date & $\overline{\rm S/N}$ &  RV & Julian Date & $\overline{\rm S/N}$ &  RV & Julian Date & $\overline{\rm S/N}$ &  RV       \\ 
(days)-2456000 & & (km/s) & (days)-2456000 & & (km/s) & (days)-2456000 & & (km/s) \\ \hline
           91.564471  &        9.5  &            $-29.153^{+0.081}_{-0.081}$ &          110.542846  &        7.5  &              $-34.89^{+0.12}_{-0.093}$ &          124.524128  &        6.5  &              $-34.625^{+0.075}_{-0.069}$ \\
           91.581828  &       13.9  &            $-29.146^{+0.085}_{-0.082}$ &          110.560305  &       11.0  &              $-34.77^{+0.11}_{-0.061}$ &          129.435272  &       10.8  &              $-33.343^{+0.081}_{-0.085}$ \\
           91.599185  &        7.0  &            $-29.397^{+0.082}_{-0.073}$ &          110.577764  &        6.4  &              $-35.15^{+0.11}_{-0.075}$ &          130.540005  &       10.0  &              $-33.237^{+0.076}_{-0.060}$ \\
          100.433076  &        8.2  &            $-24.792^{+0.073}_{-0.067}$ &          124.453660  &        6.9  &            $-34.620^{+0.093}_{-0.081}$ &          130.574986  &       18.8  &               $-33.098^{+0.077}_{-0.11}$ \\
          100.468077  &       16.2  &            $-24.749^{+0.060}_{-0.066}$ &          124.488780  &        8.2  &            $-34.500^{+0.061}_{-0.076}$ &          130.575012  &       11.1  &              $-33.291^{+0.077}_{-0.068}$ \\
          100.468144  &        9.7  &             $-24.886^{+0.067}_{-0.12}$ &          124.488856  &       13.1  &            $-34.515^{+0.068}_{-0.077}$ &          130.609940  &       10.6  &              $-33.159^{+0.069}_{-0.077}$ \\
          100.503013  &        9.5  &             $-24.834^{+0.066}_{-0.11}$ &                      &             &                                        &                      &             &                                          \\
\hline 
\end{tabular}
%\tablefoot{
%\tablefoottext{a}{}
%}
\end{table*} 

\begin{table*}[htbf]
\setlength{\extrarowheight}{2pt}
\scriptsize
\caption{KOI-0972: measured radial velocities}
\label{tab:KOI-0972}
\centering 
\begin{tabular}{ccc|ccc|ccc}     % 7 columns 
\hline\hline
 Julian Date & $\overline{\rm S/N}$ &  RV & Julian Date & $\overline{\rm S/N}$ &  RV & Julian Date & $\overline{\rm S/N}$ &  RV       \\ 
(days)-2456000 & & (km/s) & (days)-2456000 & & (km/s) & (days)-2456000 & & (km/s) \\ \hline
          100.376405  &       58.2  &                   $-8.2^{+4.2}_{-3.4}$ &          114.503592  &       53.2  &                   $-8.5^{+8.7}_{-4.4}$ &          126.348892  &       37.8  &                     $-8.8^{+3.3}_{-3.5}$ \\
          100.388716  &       86.3  &                  $-12.9^{+3.4}_{-2.6}$ &          114.651427  &       40.6  &                  $-10.0^{+3.2}_{-5.5}$ &          127.358229  &       47.4  &                    $-11.2^{+4.6}_{-5.1}$ \\
          100.401026  &       64.0  &                  $-15.4^{+3.0}_{-4.0}$ &          115.359156  &       38.5  &                  $-11.3^{+5.2}_{-4.3}$ &          128.359248  &       34.8  &                    $-14.9^{+8.8}_{-4.7}$ \\
          101.610586  &       41.2  &                  $-12.8^{+4.5}_{-4.8}$ &          115.507564  &       44.8  &                  $-14.1^{+3.8}_{-5.1}$ &          129.361016  &       37.8  &                    $-15.5^{+3.3}_{-3.7}$ \\
          101.622675  &       61.7  &                  $-11.7^{+4.5}_{-5.8}$ &          115.655973  &       22.7  &                  $-10.7^{+5.5}_{-4.1}$ &          130.359776  &       54.4  &                    $-16.1^{+4.6}_{-3.3}$ \\
          101.634765  &       45.9  &                  $-15.4^{+5.1}_{-2.5}$ &          116.430815  &       41.5  &                  $-17.0^{+4.1}_{-3.0}$ &          131.358377  &       41.9  &                    $-14.1^{+3.8}_{-4.4}$ \\
          102.370604  &       61.3  &                   $-18.8^{+6.3}_{-11}$ &          117.360460  &       34.2  &                  $-15.8^{+5.1}_{-2.7}$ &          132.358859  &       40.2  &                         $-4^{+11}_{-14}$ \\
          102.382723  &       89.3  &                   $-9.3^{+2.3}_{-4.8}$ &          117.399590  &       41.8  &                  $-13.1^{+4.9}_{-5.3}$ &          133.361742  &       46.1  &                   $-14.0^{+4.5}_{-0.92}$ \\
          102.394841  &       65.1  &                   $-7.7^{+2.7}_{-4.5}$ &          117.438719  &       24.0  &                  $-14.9^{+6.1}_{-3.0}$ &          137.356202  &       41.2  &                     $-22.5^{+6.5}_{-16}$ \\
          103.602984  &       62.3  &                  $-13.2^{+8.8}_{-5.1}$ &          118.353116  &       34.0  &                  $-18.9^{+5.4}_{-2.9}$ &          138.359861  &       43.1  &                     $-19.0^{+5.4}_{-15}$ \\
          103.615092  &       87.4  &                  $-16.4^{+6.6}_{-4.8}$ &          119.361354  &       23.1  &                  $-18.7^{+5.7}_{-3.0}$ &          140.360137  &       37.0  &                     $-7.7^{+5.0}_{-2.8}$ \\
          103.627200  &       61.2  &                  $-10.7^{+5.7}_{-4.8}$ &          120.384560  &       29.7  &                  $-22.0^{+4.6}_{-3.7}$ &          141.354705  &       39.0  &                    $-16.0^{+4.2}_{-3.6}$ \\
          110.363523  &       34.3  &                  $-13.4^{+6.5}_{-4.2}$ &          121.363682  &       40.7  &                  $-16.2^{+3.5}_{-8.6}$ &          418.660632  &       37.0  &                    $-15.0^{+4.7}_{-1.9}$ \\
          111.366741  &       39.7  &                  $-15.0^{+5.2}_{-2.5}$ &          122.427018  &       41.9  &                  $-13.0^{+3.4}_{-3.1}$ &          420.646028  &       15.4  &                       $-27^{+20}_{-5.9}$ \\
          112.358116  &       48.0  &                  $-12.2^{+5.5}_{-4.6}$ &          123.355871  &       30.5  &                   $-6.7^{+6.6}_{-4.1}$ &          420.658684  &       21.2  &                   $-82.22^{+0.97}_{-37}$ \\
          113.357987  &       34.7  &                   $-6.4^{+5.9}_{-4.8}$ &          124.352637  &       41.6  &                  $-14.2^{+3.9}_{-3.3}$ &          426.552329  &       13.7  &                         $6^{+16}_{-3.1}$ \\
          114.355757  &       34.3  &                  $-11.8^{+5.3}_{-3.7}$ &          125.446545  &       41.2  &                  $-12.2^{+3.5}_{-8.5}$ &          435.646207  &       34.0  &                         $4^{+12}_{-4.2}$ \\
\hline 
\end{tabular}
%\tablefoot{
%\tablefoottext{a}{}
%}
\end{table*} 

\begin{table*}[htbf]
\setlength{\extrarowheight}{2pt}
\scriptsize
\caption{KOI-1463: measured radial velocities}
\label{tab:KOI-1463}
\centering 
\begin{tabular}{ccc|ccc|ccc}     % 7 columns 
\hline\hline
 Julian Date & $\overline{\rm S/N}$ &  RV & Julian Date & $\overline{\rm S/N}$ &  RV & Julian Date & $\overline{\rm S/N}$ &  RV       \\ 
(days)-2456000 & & (km/s) & (days)-2456000 & & (km/s) & (days)-2456000 & & (km/s) \\ \hline
          114.381820  &       12.8  &               $-30.76^{+0.47}_{-0.47}$ &          517.458750  &       13.2  &                $31.76^{+0.28}_{-0.41}$ &          838.477732  &       12.2  &                 $306.20^{+0.33}_{-0.39}$ \\
          114.399160  &       19.8  &               $-30.93^{+0.39}_{-0.36}$ &          805.527896  &        7.3  &               $305.37^{+0.34}_{-0.34}$ &          838.494471  &       16.1  &                 $305.52^{+0.29}_{-0.19}$ \\
          114.416499  &       15.2  &               $-31.11^{+0.36}_{-0.17}$ &          805.560558  &        3.2  &               $306.09^{+0.41}_{-0.33}$ &          838.511211  &        8.9  &                 $305.54^{+0.31}_{-0.34}$ \\
          517.378382  &       14.5  &                $31.18^{+0.17}_{-0.28}$ &          805.560620  &       11.3  &               $305.58^{+0.34}_{-0.31}$ &          841.587718  &        6.9  &                 $305.43^{+0.29}_{-0.29}$ \\
          517.418566  &       20.1  &                $31.36^{+0.19}_{-0.34}$ &          805.593404  &        7.5  &               $305.57^{+0.34}_{-0.29}$ &                      &             &                                          \\
\hline 
\end{tabular}
%\tablefoot{
%\tablefoottext{a}{}
%}
\end{table*} 

\begin{table*}[htbf]
\setlength{\extrarowheight}{2pt}
\scriptsize
\caption{KOI-3725: measured radial velocities}
\label{tab:KOI-3725}
\centering 
\begin{tabular}{ccc|ccc|ccc}     % 7 columns 
\hline\hline
 Julian Date & $\overline{\rm S/N}$ &  RV & Julian Date & $\overline{\rm S/N}$ &  RV & Julian Date & $\overline{\rm S/N}$ &  RV       \\ 
(days)-2456000 & & (km/s) & (days)-2456000 & & (km/s) & (days)-2456000 & & (km/s) \\ \hline
          500.625109  &       14.2  &                  $-51.3^{+3.9}_{-2.1}$ &          504.571903  &       29.6  &                 $-30.9^{+1.2}_{-0.97}$ &          841.530484  &       14.6  &                    $-15.0^{+2.1}_{-1.5}$ \\
          501.383460  &       19.2  &                  $-46.2^{+2.0}_{-1.4}$ &          504.609225  &       22.8  &                 $-41.9^{+1.1}_{-0.73}$ &          841.536254  &       20.6  &                    $-29.0^{+2.7}_{-1.1}$ \\
          503.425012  &       23.2  &                  $-42.1^{+1.4}_{-1.3}$ &          517.650133  &       21.0  &                $-45.20^{+0.79}_{-1.3}$ &          841.542025  &       14.2  &                    $-29.1^{+1.4}_{-2.9}$ \\
          504.534581  &       19.0  &                  $-50.2^{+1.4}_{-1.2}$ &                      &             &                                        &                      &             &                                          \\
\hline 
\end{tabular}
%\tablefoot{
%\tablefoottext{a}{}
%}
\end{table*} 

\begin{table*}[htbf]
\setlength{\extrarowheight}{2pt}
\scriptsize
\caption{KOI-3728: measured radial velocities}
\label{tab:KOI-3728}
\centering 
\begin{tabular}{ccc|ccc|ccc}     % 7 columns 
\hline\hline
 Julian Date & $\overline{\rm S/N}$ &  RV & Julian Date & $\overline{\rm S/N}$ &  RV & Julian Date & $\overline{\rm S/N}$ &  RV       \\ 
(days)-2456000 & & (km/s) & (days)-2456000 & & (km/s) & (days)-2456000 & & (km/s) \\ \hline
          354.369018  &       10.1  &                  $-36.3^{+5.3}_{-3.4}$ &          511.418066  &        9.2  &                     $-44^{+10}_{-3.6}$ &          523.540289  &       15.3  &                    $-48.6^{+3.2}_{-1.9}$ \\
          354.379993  &       14.7  &                  $-37.0^{+3.8}_{-2.5}$ &          512.472273  &        6.9  &                     $-54^{+11}_{-2.5}$ &          525.518022  &       14.4  &                    $-47.9^{+4.7}_{-2.7}$ \\
          354.390968  &       10.1  &                   $-34.5^{+5.3}_{-16}$ &          513.408866  &       11.6  &                  $-48.6^{+5.6}_{-3.0}$ &          525.536150  &       20.8  &                     $-55.7^{+4.6}_{-26}$ \\
          498.570987  &       12.2  &                   $-57.8^{+4.4}_{-25}$ &          513.427128  &       18.5  &                  $-50.7^{+5.0}_{-2.2}$ &          525.554278  &       14.8  &                    $-53.9^{+5.4}_{-2.6}$ \\
          498.583268  &       16.9  &                   $-58.4^{+3.1}_{-15}$ &          513.445391  &       13.7  &                  $-50.9^{+6.0}_{-6.9}$ &          528.567258  &       11.1  &                     $-45.1^{+4.2}_{-11}$ \\
          498.595550  &       11.5  &                   $-58.3^{+2.9}_{-17}$ &          514.605639  &       14.4  &                  $-56.9^{+5.2}_{-4.3}$ &          559.339189  &       16.7  &                    $-57.0^{+3.7}_{-2.5}$ \\
          500.576958  &        4.8  &                      $-30^{+20}_{-35}$ &          514.623762  &       20.1  &                     $-49^{+10}_{-3.3}$ &          559.371563  &       15.8  &                    $-57.1^{+2.0}_{-3.9}$ \\
          500.602103  &        6.8  &                     $-67^{+25}_{-3.6}$ &          514.641885  &       14.0  &                  $-59.8^{+7.4}_{-4.5}$ &          559.403933  &       14.8  &                    $-53.7^{+5.5}_{-6.7}$ \\
          501.428000  &        3.9  &                     $-49^{+17}_{-8.2}$ &          517.590081  &       14.0  &                  $-45.7^{+4.9}_{-6.4}$ &          567.330058  &       14.7  &                    $-50.1^{+2.6}_{-4.8}$ \\
          510.446736  &        6.1  &                     $-51^{+19}_{-9.7}$ &          517.606266  &       18.5  &                  $-48.7^{+5.6}_{-3.1}$ &          567.367052  &       15.0  &                    $-47.4^{+2.5}_{-4.0}$ \\
          510.468900  &        8.9  &                     $-51^{+15}_{-4.4}$ &          517.622450  &       11.4  &                  $-50.4^{+4.0}_{-1.4}$ &          567.403480  &       14.6  &                    $-50.2^{+3.2}_{-3.8}$ \\
          510.491065  &        6.8  &                     $-24^{+28}_{-5.7}$ &          523.503712  &       13.5  &                  $-47.3^{+3.7}_{-5.6}$ &          597.418633  &       12.6  &                       $-48^{+25}_{-3.2}$ \\
          511.390550  &       11.0  &                  $-53.3^{+4.5}_{-4.9}$ &          523.522001  &       20.9  &                  $-49.4^{+4.7}_{-2.2}$ &          597.418633  &        9.3  &                        $-47^{+14}_{-76}$ \\
          511.404308  &       14.7  &                  $-56.3^{+4.0}_{-5.7}$ &                      &             &                                        &                      &             &                                          \\
\hline 
\end{tabular}
%\tablefoot{
%\tablefoottext{a}{}
%}
\end{table*} 

\begin{table*}[htbf]
\setlength{\extrarowheight}{2pt}
\scriptsize
\caption{KOI-3853: measured radial velocities}
\label{tab:KOI-3853}
\centering 
\begin{tabular}{ccc|ccc|ccc}     % 7 columns 
\hline\hline
 Julian Date & $\overline{\rm S/N}$ &  RV & Julian Date & $\overline{\rm S/N}$ &  RV & Julian Date & $\overline{\rm S/N}$ &  RV       \\ 
(days)-2456000 & & (km/s) & (days)-2456000 & & (km/s) & (days)-2456000 & & (km/s) \\ \hline
          498.550774  &       19.3  &            $-88.037^{+0.040}_{-0.040}$ &          511.624458  &       11.8  &            $-57.982^{+0.054}_{-0.070}$ &          528.494410  &       28.8  &              $-59.539^{+0.052}_{-0.049}$ \\
          500.427256  &        6.8  &            $-78.182^{+0.051}_{-0.051}$ &          511.656257  &        9.6  &            $-57.901^{+0.070}_{-0.057}$ &          528.515065  &       43.6  &              $-59.420^{+0.053}_{-0.047}$ \\
          503.522251  &       23.8  &            $-67.622^{+0.057}_{-0.057}$ &          513.591688  &       26.4  &            $-70.695^{+0.057}_{-0.058}$ &          528.535719  &       32.7  &              $-59.298^{+0.054}_{-0.053}$ \\
          504.628523  &       24.9  &            $-64.491^{+0.053}_{-0.053}$ &          515.538671  &       25.8  &           $-101.989^{+0.058}_{-0.053}$ &          559.438923  &       28.5  &             $-108.981^{+0.049}_{-0.057}$ \\
          505.425120  &       25.4  &            $-62.606^{+0.048}_{-0.048}$ &          522.559858  &       24.5  &            $-76.193^{+0.053}_{-0.050}$ &          559.449900  &       39.6  &             $-108.981^{+0.048}_{-0.056}$ \\
          510.419882  &       21.0  &            $-56.134^{+0.052}_{-0.052}$ &          526.420908  &       35.1  &            $-64.086^{+0.050}_{-0.048}$ &          559.460877  &       27.4  &             $-109.009^{+0.047}_{-0.048}$ \\
          511.363906  &       25.3  &            $-57.187^{+0.051}_{-0.051}$ &          526.439128  &       45.2  &            $-64.033^{+0.048}_{-0.052}$ &          569.430575  &       25.6  &              $-64.360^{+0.053}_{-0.053}$ \\
          511.553563  &       33.1  &            $-57.133^{+0.056}_{-0.056}$ &          526.457348  &       27.8  &            $-63.921^{+0.048}_{-0.054}$ &          569.467568  &       16.5  &              $-64.290^{+0.057}_{-0.048}$ \\
          511.569631  &       14.5  &            $-57.611^{+0.056}_{-0.054}$ &                      &             &                                        &                      &             &                                          \\
\hline 
\end{tabular}
%\tablefoot{
%\tablefoottext{a}{}
%}
\end{table*} 

\begin{table*}[htbf]
\setlength{\extrarowheight}{2pt}
\scriptsize
\caption{KOI-3890: measured radial velocities}
\label{tab:KOI-3890}
\centering 
\begin{tabular}{ccc|ccc|ccc}     % 7 columns 
\hline\hline
 Julian Date & $\overline{\rm S/N}$ &  RV & Julian Date & $\overline{\rm S/N}$ &  RV & Julian Date & $\overline{\rm S/N}$ &  RV       \\ 
(days)-2456000 & & (km/s) & (days)-2456000 & & (km/s) & (days)-2456000 & & (km/s) \\ \hline
          499.397905  &        6.7  &            $-27.368^{+0.087}_{-0.087}$ &          511.496364  &        2.8  &              $-26.35^{+0.16}_{-0.074}$ &          516.552987  &        9.4  &               $-25.955^{+0.073}_{-0.38}$ \\
          499.438485  &        4.1  &               $-27.74^{+0.19}_{-0.19}$ &          511.535899  &        5.5  &            $-26.176^{+0.074}_{-0.052}$ &          516.554295  &       16.9  &              $-25.835^{+0.064}_{-0.098}$ \\
          502.629346  &        6.2  &            $-27.330^{+0.072}_{-0.072}$ &          513.484972  &        7.6  &            $-26.418^{+0.052}_{-0.072}$ &          516.589283  &       10.0  &              $-25.870^{+0.066}_{-0.070}$ \\
          510.526583  &        5.0  &              $-26.60^{+0.10}_{-0.100}$ &          513.522807  &       14.2  &            $-26.240^{+0.055}_{-0.056}$ &          522.447180  &        3.2  &                $-24.71^{+0.38}_{-0.090}$ \\
          510.564787  &        2.5  &               $-26.35^{+0.24}_{-0.24}$ &          513.523404  &        8.2  &            $-26.140^{+0.072}_{-0.065}$ &          522.485030  &       12.2  &              $-24.798^{+0.096}_{-0.055}$ \\
          510.577098  &        8.2  &            $-26.391^{+0.090}_{-0.074}$ &          513.560045  &        8.5  &            $-26.260^{+0.056}_{-0.073}$ &          522.485601  &        7.2  &              $-24.970^{+0.098}_{-0.064}$ \\
          510.639922  &        7.1  &            $-26.387^{+0.074}_{-0.082}$ &          516.520616  &        9.8  &            $-25.942^{+0.065}_{-0.066}$ &          522.522308  &        8.8  &              $-25.041^{+0.070}_{-0.096}$ \\
          511.457622  &        8.2  &             $-26.370^{+0.082}_{-0.16}$ &                      &             &                                        &                      &             &                                          \\
\hline 
\end{tabular}
%\tablefoot{
%\tablefoottext{a}{}
%}
\end{table*}

\end{document}